\documentclass[twocolumn]{aastex701}
\usepackage[T1]{fontenc}
\usepackage{multirow} 
\usepackage{longtable}
\usepackage{newtxtext,newtxmath}
\usepackage{amsmath}	% Advanced maths commands
\usepackage{upgreek}

\begin{document}

\title{Evidence for Systematically Larger Dust Grains in Upper Scorpius Relative to Taurus Disks}

\author[0000-0002-7616-666X]{Yao Liu}
\affiliation{School of Physical Science and Technology, Southwest Jiaotong University, Chengdu 610031, China}
\email[show]{yliu@swjtu.edu.cn}  

\author[0000-0001-7962-1683]{Ilaria Pascucci}
\affiliation{Lunar and Planetary Laboratory, The University of Arizona, Tucson, AZ 85721, USA}
\email{fakeemail3@google.com}

\author{Fei Gao}
\affiliation{School of Physical Science and Technology, Southwest Jiaotong University, Chengdu 610031, China}
\email{fakeemail3@google.com}

\author[0000-0001-8184-5547]{Chengyan Xie}
\affiliation{Lunar and Planetary Laboratory, The University of Arizona, Tucson, AZ 85721, USA}
\email{fakeemail3@google.com}

\author[0000-0002-7607-719X]{Feng Long}
\affiliation{Kavli Institute for Astronomy and Astrophysics, Peking University, Beijing 100871, China}
\email{fakeemail3@google.com}

\author[0000-0003-2251-0602]{John Carpenter}
\affiliation{Joint ALMA Observatory, Avenida Alonso de C{\' o}rdova 3107, Vitacura, Santiago, Chile}
\email{fakeemail3@google.com}

\author[0000-0001-7552-1562]{Klaus M. Pontoppidan}
\affiliation{Jet Propulsion Laboratory, California Institute of Technology, 4800 Oak Grove Drive, Pasadena, CA, 91109, USA}
\email{fakeemail3@google.com}

\author[0000-0003-4335-0900]{Andrea Banzatti}
\affiliation{Department of Physics, Texas State University, 749 North Comanche Street, San Marcos, TX 78666, USA}
\email{fakeemail3@google.com}

\author[0000-0002-0364-937X]{Richard Booth}
\affiliation{School of Physics and Astronomy, University of Leeds, Leeds, LS2 9JT, UK}
\email{fakeemail3@google.com}

\author[0000-0002-2314-7289]{Steve Ertel}
\affiliation{Department of Astronomy and Steward Observatory, University of Arizona, 933 N Cherry Ave., Tucson, AZ 85721-0065, USA}
\affiliation{Large Binocular Telescope Observatory, University of Arizona, 933 N Cherry Ave., Tucson, AZ 85721-0065, USA}
\email{fakeemail3@google.com}

\author[0000-0001-8060-1321]{Min Fang}
\affiliation{Purple Mountain Observatory, Chinese Academy of Sciences, 10 Yuanhua Road, Nanjing 210023, People’s Republic of China}
\affiliation{University of Science and Technology of China, Hefei 230026, People’s Republic of China}
\email{fakeemail3@google.com}

\author{Uma Gorti}
\affiliation{NASA Ames Research Center, Moffett Field, CA 94035, USA}
\affiliation{Carl Sagan Center, SETI Institute, Mountain View, CA 94043, USA}
\email{fakeemail3@google.com}

\author[0000-0003-0448-6354]{Tamara Molyarova}
\affiliation{School of Physics and Astronomy, University of Leeds, Leeds, LS2 9JT, UK}
\email{fakeemail3@google.com}

%\author[0009-0002-2380-6683]{Eshan Raul}
%\affiliation{Department of Astronomy, University of Wisconsin--Madison, Madison, WI 53706, USA}
%\email{fakeemail3@google.com}

%% \collaboration{all}{The Terra Mater collaboration}

\begin{abstract}

Infrared spectroscopy provides a powerful diagnostic for probing the mineralogical properties of dust grains in the terrestrial planet–forming regions of protoplanetary disks. The Upper Scorpius association offers an excellent laboratory for studying disk evolution because it represents an evolved stage ($5{-}10\,{\rm Myr}$) compared with younger star-forming regions such as the Taurus Molecular Cloud ($1{-}3\,{\rm Myr}$). In this work, we present mid-infrared spectra of 11 disks in Upper Scorpius that were obtained with the Mid-Infrared Instrument aboard the James Webb Space Telescope. We derive emission feature indices for crystalline olivine and pyroxene centered at ${\sim}\,9.2\,\mu{\rm m}$ and ${\sim}\,11.1\,\mu{\rm m}$, as well as perform spectral decomposition to quantify dust crystallinity and characteristic grain size. These results are compared with those measured from Spitzer/IRS spectra of 31 disks in Taurus with similar stellar types. We find no significant difference in dust crystallinity between the two groups, suggesting that crystallization is largely established at early stages of disk evolution. Our analysis indicates that the average grain size in Upper Scorpius disks is systematically larger than that in Taurus disks, aligning with theories of dust evolution. We also observe a trend of increasing grain size towards later-type stars, as well as a correlation between crystallinity, grain size and the flux ratio $F_{24}/F_{8}$, which serves as a measure of dust settling. These results suggest that dust processing proceeds in tandem with disk evolution.

\end{abstract}


\section{Introduction} 

Protoplanetary disks form as part of the star-formation process, and contain abundant gas and dust grains that are essential for planet formation \citep[e.g.,][]{Williams2011,2022Raymond,Drazkowska2023}. During the early stages of disk's evolution, frequent collisions occur between submicron particles due to processes such as Brownian motion and turbulence particularly where densities are high. If these collisions are not violent, the particles adhere to each other via electrostatic forces. Once the particles overcome growth obstacles like bouncing and fragmentation and are sufficiently concentrated in pressure maxima, streaming instability can lead to the formation of km-size bodies and planetesimals  \citep[e.g.,][]{2005Dullemond,Liu2020,2022Raymond,Birnstiel2024}. Through processes like pebble accretion \citep[e.g.,][]{2010Ormel}, planetesimals gradually turn into protoplanets. In the inner disk region, low-mass planetary embryos evolve into terrestrial planets, while in the outer disk more massive planetary embryos can accrete a large amount of gas to form giant planets \citep{1993Lissauer,2022Raymond}. Celestial bodies within the solar system, such as comets and asteroids, contain clues about the early formation of our solar system. Similarly, analyzing dust in protoplanetary disks provides insights into the early stages of planet formation. 

Silicates are the most common and dominant component of dust grains in protoplanetary disks \citep{Henning2010}. Under certain conditions, such as high temperatures caused by viscous heating \citep{Gail2004}, accretion outbursts \citep{Abraham2009,Lee2026}, and/or shocks \citep{2005ASPC..341..849D}, crystalline silicates can form through annealing or gas-phase condensation. Crystalline silicates such as forsterite (${\rm Mg_2SiO_4}$) and enstatite (${\rm MgSiO_3}$) have several strong and narrow resonances around 10\,$\mu {\rm m}$. The crystalline mass fraction of the interstellar medium dust near the center of the Milky Way is ${\sim}\,1\%$ \citep[e.g.,][]{2004Kemper,2016Fogerty}. However, the crystallinity of cometary dust approaches ${\sim}\,20\%$ and can be as high as ${\sim}\,55\%$ \citep[e.g.,][]{2021Woodward, 2023PSJ.....4..242H}. It is believed that these crystalline silicates originally formed in high-temperature regions of the solar nebula, and were transported to cold outer regions through mechanisms like disk winds or turbulence \citep{Giacalone2019,Ciesla2011}. Amorphous silicates exhibit broad and smooth features at 10\,$\mu {\rm m}$ and 18\,$\mu {\rm m}$ attributed to the Si-O stretching vibrational mode and O-Si-O bending vibrational mode, respectively.  Therefore, infrared spectroscopy is a powerful tool for diagnosing dust mineralogy and uncovering the thermal and dynamical history of protoplanetary disks.

Owing largely to the Spitzer Space Telescope, infrared spectroscopic observations have been carried out for a large sample of disks with ages of ${\sim}\,1\,{-}\,3$\,Myr. For instance, \citet{Furlan2011} presented the InfraRed Spectrograph (IRS) spectra for 161 T Tauri stars and young brown dwarfs in the Taurus molecular cloud. \citet{Manoj2011} carried out a detailed analysis of the IRS spectra of 68 Class II objects in the Chamaeleon I star-forming region. \citet{McClure2010} analyzed the IRS spectra of 136 young stellar objects in the Ophiuchus cloud. \citet{Juhasz2010} characterized the constituents of protoplanetary dust around 45 Herbig Ae/Be stars. Collectively, these extensive observations and analysis have revealed that most disks exhibit significant dust processing, including grain growth, crystallization, and substantial vertical settling, even at young ages of $1\,{-}\,3$\,Myr.  Nevertheless, spectroscopic observations at infrared wavelengths for stars at older ages are quite limited. The Upper Scorpius association (hereafter USco) provides an ideal target for such studies. The 5–10 Myr age of this association \citep{Preibisch2002,Pecaut2012,David2019} implies that their protoplanetary disks are in the last stage of evolution before dissipation \citep{Hernandez2008,Williams2011}.

In this work, we present mid-infrared spectra of 11 disks in USco obtained with the Mid-Infrared Instrument (MIRI) on the James Webb Space Telescope (JWST), and investigate the mineralogical properties of dust grains in these evolved systems. Observations and data reduction are described in Sect.~\ref{sec:obs}. In Sect.~\ref{sec:cryidx}, we compare the crystalline feature indices of USco disks to those of a sample of Taurus disks to study dust processing as a function of time. Sect.~\ref{sec:analysis} focuses on a detailed decomposition of the $10\,\mu{\rm m}$ silicate feature for both the USco and Taurus disks, with the goal of quantitatively constraining the dust crystallinity and grain size. We briefly discuss crystalline emission features shown in the long wavelength regime in Sect~\ref{sec:longwav}. The paper closes with a brief summary in Sect.~\ref{sec:summary}.

\begin{figure*}
\centering
\includegraphics[width=0.9\textwidth, angle=0]{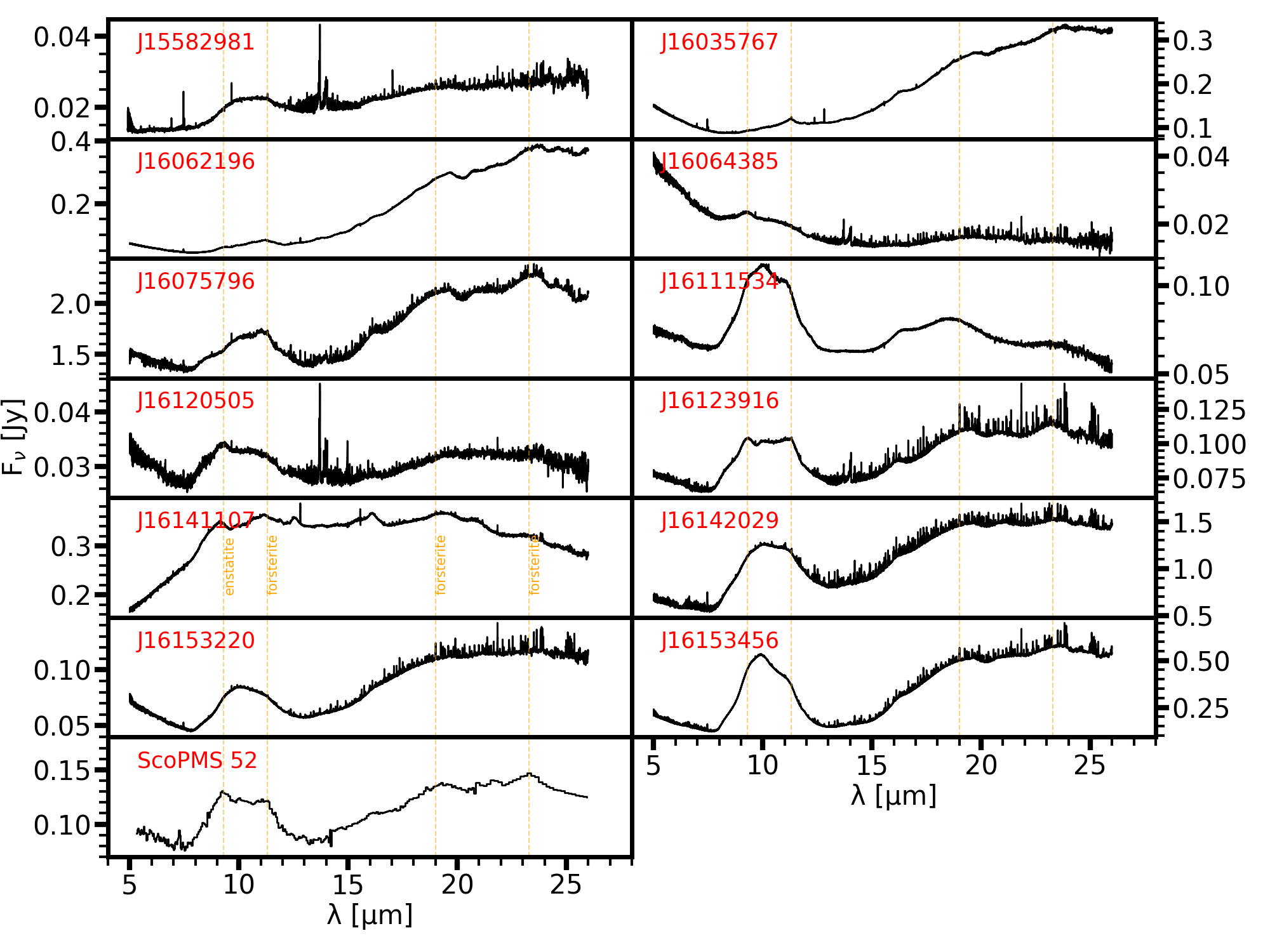}
\caption{Mid-infrared spectra of USco disks studied in this work. Most of the spectra are from the MIRI/MRS program 2970 \citep{Pascucci2023jwst}. The MIRI spectrum of J15582981 is from the MINDS program \citep{Henning2024}. For ScoPMS\,52, we show the IRS data obtained from the FEPS Spitzer Legacy program \citep{Meyer2006}. The wavelengths of the most prominent crystalline dust features are indicated by the vertical dashed lines.}
\label{fig:obs}
\end{figure*}

\begin{table*}
\centering
\caption{Properties of the targets studied in this work.}
\label{tab:target}
\begin{tabular}{ccccccccc}
\hline
2MASS ID  &  SpT   &  $M_{\star}$ [$M_{\odot}$]    &  ${\rm log}_{10}\left(\frac{L_{\star}}{L_{\odot}}\right)$ &  ${\rm log}_{10}\left(\frac{\dot{M}_{\rm acc}}{M_{\odot}/{\rm yr}}\right)$  &  $D$ [pc] & $F_{\rm mm}$ [mJy] &  Reference  &  Instrument \\
\hline
\multicolumn{9}{c}{Old USco sample}\\
\hline   
J15582981-2310077  &    M4.8     &   0.11    &    $-1.39$   & ${-}8.92$     &  141.1 & 5.93 & 1 & MIRI \\
J16035767-2031055  &    K5.1     &   0.81    &    $-0.20$   & ${-}9.29$     &  142.6 & 5.83 & 1 & MIRI \\
J16062196-1928445  &    M0.9     &   0.42    &    $-0.45$   & ${-}8.90$     &  142.0 & 4.87 & 1 & MIRI \\
J16064385-1908056  &    K7.9     &   0.69    &    $-0.47$   & ${-}9.95$     &  145.3 & $<0.75$ & 1 & MIRI \\
J16075796-2040087  &    K4.0     &   0.71    &    $-0.96$   & ${-}8.96$     &  135.9 & 23.9 & 1 & MIRI \\
J16111534-1757214  &    M1.2     &   0.40    &  $-0.55$  & ${<}\,{-}9.69$   &  135.3 & $<0.65$ & 1 & MIRI \\
J16120505-2043404  &    M1.5     &   0.37    &    $-0.57$   & ${-}9.50$     &  122.5 & 3.47 & 1 & MIRI \\
J16123916-1859284  &    M2.0     &   0.33    &    $-0.54$   & ${-}8.69$     &  134.7 & 9.45 & 1 & MIRI \\
J16141107-2305362  &    K3.0     &   1.23    &    $0.50$    &  $-$          &  142.0 & 5.05 & 1 & MIRI \\
J16142029-1906481  &    K9.0     &   0.67    &    $-0.68$   & ${-}9.04$     &  138.8 & 41.45 & 1 & MIRI \\
J16153220-2010236  &    M2.4     &   0.31    &    $-0.57$   & ${-}9.13$     &  142.0 & 1.92 & 1 & MIRI \\
J16153456-2242421  &    M0.2     &   0.48    &    $-0.41$   & ${-}8.68$     &  136.9 & 12.84 & 1 & MIRI \\
ScoPMS\,52         &    K6.0     &   0.55    &    $0.17$    &  $-$          &  134.5 & $<10$ & $-$ & IRS \\
\hline 
Median             &    M0.2    &    0.48    &    $-0.54$   &   $-9.09$   &    $-$     & 5.02  &  $-$  & $-$  \\
\hline
\multicolumn{9}{c}{Young Taurus sample}\\
\hline
AA Tau & K7.0  & 0.79   & $-$0.093  & $-$8.48  & 134.7 & 139.4 & 2  & IRS  \\
BP Tau & K7.0  & 0.79   & $-$0.094  & $-$7.54  & 127.4 & 129.7 & 2  & IRS \\
CI Tau & K7.0  & 0.79   & 0.086     & $-$7.19  & 160.3 & 263.6 & 2  & IRS \\
CW Tau & K3.0  & 1.58   & 0.330     & $-$7.99  & 131.6 & 160.1 & 3  & IRS  \\
CX Tau & M2.5  & 0.36   & $-$0.518  & $-$8.97  & 126.7 & 25.1 & 2  & IRS  \\
CY Tau & M1.5  & 0.44   & $-$0.487  & $-$8.12  & 126.3 & 163.7 & 2  & IRS  \\
DE Tau & M1.0  & 0.39   & $-$0.064  & $-$7.58  & 128.0 & 84.3  & 2  & IRS  \\
DG Tau & K6.0  & 0.91   & 0.226     & $-$7.34  & 125.3 & 944.7 & 3   & IRS  \\
DL Tau & K7.0  & 0.81   & $-$0.015  & $-$7.17  & 159.9 & 470.2 & 3  & IRS  \\
DM Tau & M1.0  & 0.47   & $-$0.621  & $-$7.95  & 144.0 & 237.0 & 2  & IRS  \\
DN Tau & M0.0  & 0.60   & $-$0.178  & $-$8.46  & 128.6 & 210.1 & 2  & IRS  \\
DO Tau & M0.0  & 0.47   & 0.135     & $-$6.84  & 138.5 & 252.8 & 2  & IRS  \\
DR Tau & K5.0  & 1.20   & 0.538     & $-$7.50  & 193.0 & 314.9 & 3  & IRS  \\
DS Tau & K5.0  & 1.05   & $-$0.012  & $-$7.89  & 158.4 & 40.5 & 2  & IRS \\
FM Tau & M0.0  & 0.60   & $-$0.444  & $-$8.45  & 132.0 & 28.8 & 2  & IRS  \\
FN Tau & M5.0  & 0.23   & $-$0.147  & $-$10.05 & 129.9 & 36.5 & 4  & IRS  \\
FP Tau & M4.0  & 0.22   & $-$0.571  & $-$9.45  & 127.5 & $<24.3$ & 3  & IRS  \\
FT Tau & M3.0  & 0.72   & $-$0.314  & $-$8.92  & 130.2 & 121.7 & 5  & IRS  \\
FZ Tau & M0.0  & 0.45   & 0.031     & $-$7.70  & 129.2 & 30.6 & 3  & IRS  \\
GI Tau & K7.0  & 0.79   & $-$0.028  & $-$8.02  & 129.4 & 31.3 & 2  & IRS \\
GK Tau & K7.0  & 0.78   & 0.059     & $-$8.19  & 129.1 & $<13.2$  &  2 & IRS \\
GM Aur & K3.0  & 1.35   & 0.197     & $-$8.02  & 158.1 & 546.6 &  2 & IRS \\
GO Tau & M0.0  & 0.63   & $-$0.529  & $-$7.93  & 142.4 & 162.4 & 2  & IRS \\
HO Tau & M0.5  & 0.56   & $-$0.746  & $-$8.86  & 164.5 & 42.7 & 2  & IRS \\
HP Tau & K3.0  & 1.51   & 0.467     & $-$10.10 & 171.2 & 113.6 & 4  & IRS \\
IP Tau & M0.0  & 0.60   & $-$0.400  & $-$9.10  & 129.4 & 32.0  & 2  & IRS \\
IQ Tau & M0.5  & 0.54   & $-$0.143  & $-$7.55  & 131.5 & 166.8 & 2  & IRS \\
LkCa 15& K5.0  & 1.05   & 0.009     & $-$8.87  & 157.2 & 385.2 & 2  & IRS \\
V410 X-ray 1 & M4.0  &  0.25  & $-$0.409  & $-$      & 131.0 & $<9.0$ & $-$  & IRS \\
V836 Tau     & K7.0  &  0.83  & $-$0.093  & $-$8.37  & 167.0 & 67.3 & 5 & IRS \\
ZZ Tau IRS   & M5.0  &  0.12  & $-$1.266  & $-$      & 105.7 & 276.3 & $-$  & IRS  \\
\hline
Median       & M0.0  &  0.63  & $-0.093$  & $-$8.02  &  $-$  & 125.7 & $-$  &  $-$   \\
\hline
\end{tabular}
\tablecomments{References for $\dot{M}_{\rm acc}$: (1) \citet{Fang2023}; (2) \citet{Hartmann1998}; (3) \citet{White2001}; \\ (4) \citet{Lin2023}; (5) \citet{Gangi2022}}
\end{table*}

\section{Observation and data reduction} 
\label{sec:obs}

We combine data from two different MIRI/MRS programs and the Formation and Evolution of Planetary Systems (FEPS) Spitzer Legacy program to build the sample of evolved disks in USco. The majority of the sample (11 objects) is drawn from the JWST General Observer program 2970 \citep[PI: I.~Pascucci;][]{Pascucci2023jwst}, which observed 14 targets spanning a relatively narrow range in spectral type (K2–M2), corresponding to stellar masses ${\sim}\,0.3$–$1.3\,M_{\odot}$, but covering a wide range in 0.89\,mm flux densities from ${\sim}\,215\,{\rm mJy}$ to below $0.7\,\rm{mJy}$. Additional details on the sample properties, observational strategy and data reduction are provided in Xie et al. (2026, submitted). Of the 14 targets observed under PID 2970, 11 exhibit prominent silicate emission features around $10\,\mu{\rm m}$. The MIRI spectra of these 11 objects are presented in Figure~\ref{fig:obs}, and their fundamental properties are summarized in Table~\ref{tab:target}. Inspection of the MIRI Channel 1 data cubes led Xie et al. (2026, submitted) to identify four binaries: J16062196-1928445, J16120505-2043404, J16141107-2305362, and J16153220-2010236.

\begin{figure}
\centering
\includegraphics[width=0.49\textwidth, angle=0]{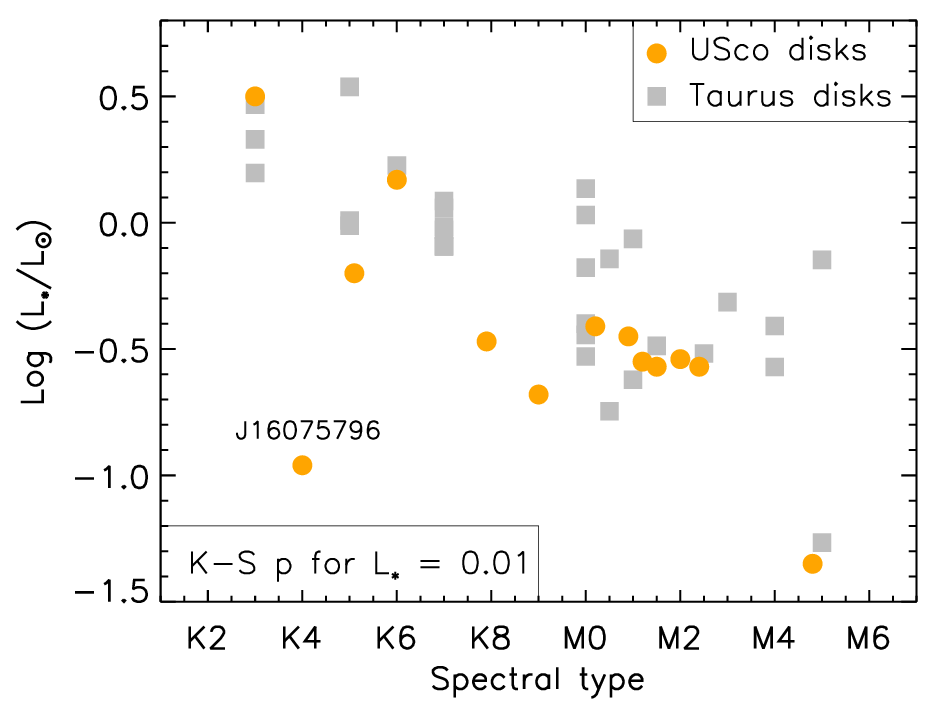}
\caption{Spectral type and stellar luminosity of the targets investigated in this work. The orange dots represent the USco sample, whereas the Taurus objects are indicated with gray squares. The K-S test p value for the distribution of stellar luminosity is low, suggesting that the two samples are different in this parameter, as expected given their different ages. Note that J16075796 has an edge-on disk, implying that the estimated stellar luminosity is quite uncertain.}
\label{fig:starprop}
\end{figure}

Two protoplanetary disks in USco were observed as part of the MIRI Mid-INfrared Disk Survey (MINDS) JWST guaranteed time program \citep[PID 1282, PI: T. Henning,][]{Henning2024}. The reduced spectra of J15582981-2310077 and J16053215-1933159 are publicly available on the MINDS website\footnote{\url{https://minds.cab.inta-csic.es/}}. We include J15582981-2310077 (M4.5) in our analysis to increase the sample size. In contrast, J16053215-1933159 does not exhibit silicate emission features near $10\,\mu{\rm m}$ \citep{Tabone2023} and is therefore excluded. The FEPS Spitzer Legacy program was designed to characterize the evolution of circumstellar gas and dust around solar-type stars \citep{Meyer2006}. This survey includes several USco disks \citep{Carpenter2008}, of which only ScoPMS\,52 and J16141107-2305362 exhibit prominent silicate emission features. The latter was also observed under our MIRI program (PID 2970). We therefore incorporate the IRS spectrum of ScoPMS\,52 into our analysis. In total, we build a sample of 13 USco disks with silicate emission features covering a range of spectral type from K3 to approximately M5, representing the old disk population

The young counterparts were directly selected from the 65 disks in the Taurus star-forming region analyzed by \citet{Sargent2009}. The IRS spectra and a homogeneous derivation for the stellar properties are available for these objects \citep{Furlan2011,Andrews2013}. To eliminate the effects of stellar properties on dust mineralogy \citep{Apai2005,Pascucci2009,Arabhavi2025,Jang2025}, we selected objects with similar spectral types to those of the USco sample. Particularly, targets with spectral types outside the range from K3 to M5 were not taken into account. Finally, we removed binary systems. The remaining 31 Taurus disks form the young disk group. 

Table~\ref{tab:target} summarizes the properties of all objects analyzed in this study. For the USco disks, we compiled the spectral type (SpT), stellar mass ($M_{\star}$), stellar luminosity ($L_{\star}$), mass accretion rate ($\dot{M}_{\rm acc}$), and 0.89\,mm flux density ($F_{\rm mm}$) from \citet{Luhman2020}, \citet{Fang2023}, and \citet{Carpenter2025}. The stellar masses span ${\sim}\,0.1$–$1.2\,M_{\odot}$, while stellar luminosities range from 0.04 to $3.16\,L_{\odot}$. The accretion rates lie between $1.1 \times 10^{-10}$ and $2.1 \times 10^{-9}\,M_{\odot}\,{\rm yr^{-1}}$, and the millimeter flux densities range from 41.5 to below 0.7\,mJy. For the Taurus disks, we adopted stellar and disk properties from \citet{Hartmann1998}, \citet{Gullbring1998}, \citet{White2001}, \citet{Andrews2013}, \citet{Gangi2022}, and \citet{Lin2023}. The stellar luminosities reported by \citet{Andrews2013} were rescaled using distances ($D$) derived from Gaia parallaxes \citep{Gaia2023}. The stellar masses span ${\sim}\,0.1$–$1.6\,M_{\odot}$, and stellar luminosities range from 0.05 to $3.45\,L_{\odot}$. The accretion rates range from $7.9 \times 10^{-11}$ to $1.4 \times 10^{-7}\,M_{\odot}\,{\rm yr^{-1}}$, while the millimeter flux densities span 945 to the upper limit of 9\,mJy. We used the Kaplan–Meier estimator implemented in the \texttt{ASURV} package to derive median values for each parameter, properly accounting for upper limits \citep{Feigelson1985, Isobe1986}. The results are provided in Table~\ref{tab:target}. Overall, Taurus disks exhibit systematically higher stellar luminosities, accretion rates, and millimeter flux densities compared to USco disks, as expected given the younger age of these systems with respect to those in USco. Figure~\ref{fig:starprop} presents the spectral types and stellar luminosities of both samples. A Kolmogorov–Smirnov (K–S) test is conducted to compare the two samples in terms of stellar luminosity. Following standard practice in astrophysics, a significance level of $\alpha\,{=}\,0.05$ (i.e., a 5\% probability of incorrectly rejecting the null hypothesis) was adopted. The resulting p value of 0.01 is smaller than this threshold, indicating that the two samples are unlikely to be drawn from the same parent population. This is expected given the older age of the USco region.
 
\section{Crystalline Dust Emission Features}
\label{sec:cryidx}

There are generally two ways to form crystalline silicates: annealing of amorphous silicates at high temperature and gas phase condensation of silicates from hot gas that is cooling \citep[e.g.,][]{Wooden2005,Petaev2005}. The inner regions of protoplanetary disks are likely to provide these conditions. For example, viscous heating in the midplane of the inner disk can produce temperatures above 1,000\,K \citep{Gail2004}. Due to the heating by shock waves or stellar outbursts, crystalline silicates can form through thermal annealing at a few AU from the central star \citep{Harker2002,Abraham2009}. Therefore, investigating the properties of crystals helps to understand the dynamic evolution of protoplanetary disks, and can also provide important insights into the early environmental conditions of the solar nebula by comparing dust properties in protoplanetary disks with those in cometary and asteroid dust. 

\begin{figure}
\centering
\includegraphics[width=0.45\textwidth, angle=0]{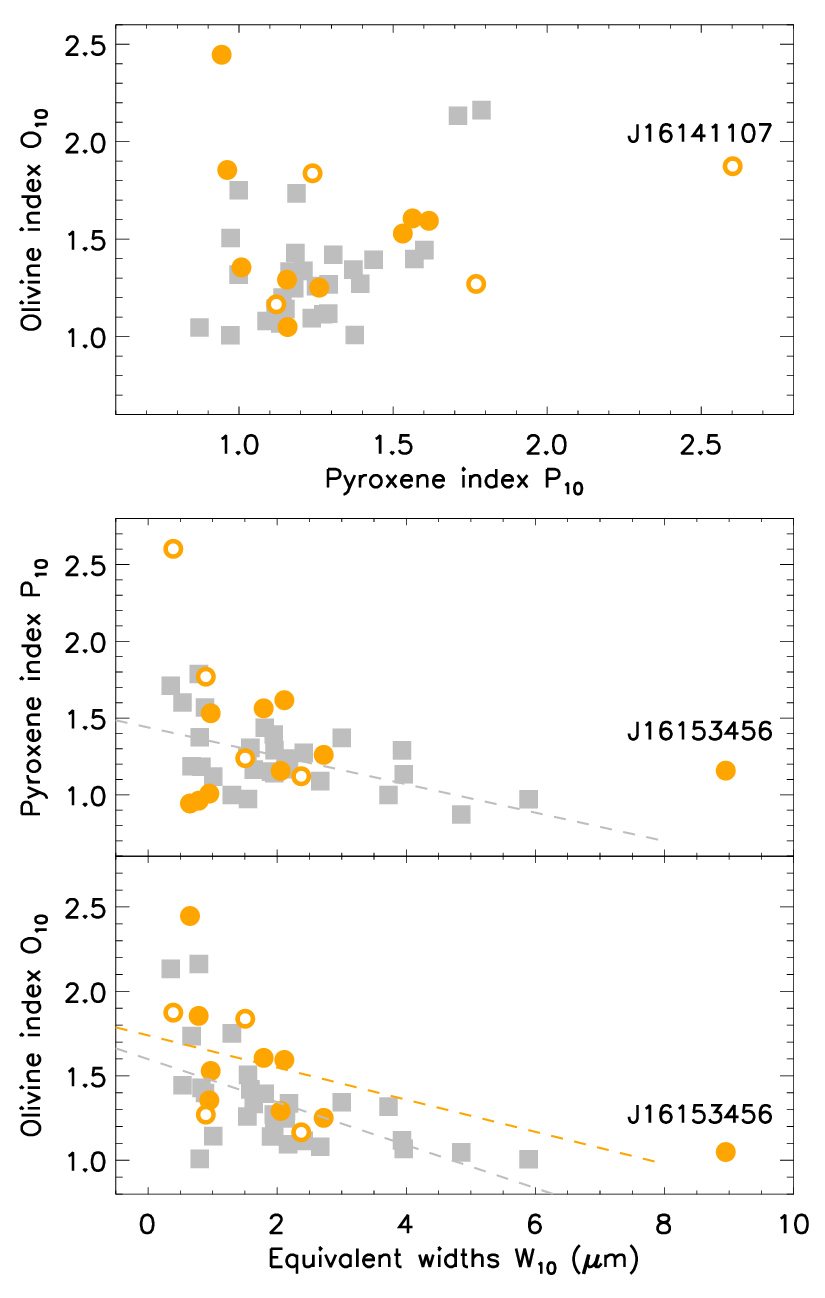}
\caption{Trends among the $10\,\mu{\rm m}$ equivalent width ($W_{10}$) and crystalline feature indices ($P_{10}$ and $O_{10}$). The Taurus disks are indicated with gray squares. Single systems in USco are represented by orange dots, while the four binaries in USco (see Sect.~\ref{sec:obs}) are marked with orange circles. We highlight two outliers: J16141107, which exhibits a particularly large pyroxene index, and J16153456, which shows an unusually large equivalent width. Table~\ref{tab:statistic} summarizes the results of K-S tests between the two samples. The Kendall’s $\tau$ and the corresponding probability $P$ values from the correlation tests for each pair of quantities are listed in Table~\ref{tab:kendall}. The dashed lines indicate statistically significant trends (Kendall’s $P\,{<}\,0.05$).}
\label{fig:index}
\end{figure}

While amorphous silicates produce a broad, smooth emission feature peaking around $10\,\mu{\rm m}$ and $19\,\mu{\rm m}$, crystalline silicates give rise to narrower, sharper substructures at characteristic wavelengths superimposed on the broader feature \citep{Henning2010}. As can be seen in Figure~\ref{fig:obs}, crystalline dust features at $9.2\,\mu{\rm m}$, $11.1\,\mu{\rm m}$ and $19\,\mu{\rm m}$ are clearly observed in most of the targets. In principle, the mass fractions of each crystalline species can be determined by fitting the spectra with disk models in combination with dust opacities of various components. Simplifications to disk models are frequently made to accelerate the parameter exploration. For instance, one would assume that the observed dust features fully originate from an optically thin layer of the disk \citep[e.g.,][]{vanBoekel2005,Bouwman2008}. Moreover, the dust temperature is not calculated self-consistently by solving the radiative transfer problem. Instead, it is usually treated as a free parameter. Some studies have adopted a power-law temperature profile \citep[e.g.,][]{Juhasz2010,Kaeufer2024,Liu2025}, while others have employed solutions from hydrostatic and radiative equilibrium models of passive disks \citep{Liu2019}. In this work, we first follow the methodology of \citet{Watson2009} to derive crystalline feature indices directly from the observed spectra, then apply spectral decomposition techniques to quantitatively constrain the crystalline mass fractions.  

\begin{table}[!t]
\caption{Median level of dust properties and the K$-$S test p values.}
\centering
\linespread{1.5}\selectfont
\label{tab:statistic}
\begin{tabular}{ccccccc}
 \hline
Sample & $N$ & $O_{10}$ & $P_{10}$ & $W_{10}$ & $C$ & $\langle a_{\rm am} \rangle$  \\
\hline
USco   & 13 & 1.53 & 1.24 & 1.51 & 17.69\% & $\mathbf{1.18\,\mu{\rm m}}$  \\
Taurus & 31 & 1.27 & 1.21 & 1.91 & 12.00\% & $\mathbf{0.16\,\mu{\rm m}}$  \\
\hline
K$-$S p & $-$ & 0.07 & 0.66 & 0.69 & 0.59 & {\bf 0.02} \\
\hline
\end{tabular}
\tablecomments{The number of objects included in the tests is denoted by $N$. The quantities $O_{10}$, $P_{10}$ and $W_{10}$ represent the olivine feature index, pyroxene feature index and equivalent width measured in the vicinity of $10\,\mu{\rm m}$, respectively, see Sect.~\ref{sec:cryidx}. The dust crystallinity and mass-averaged grain size (radius) of amorphous silicates are denoted as $C$ and $\langle a_{\rm am} \rangle$, see Sect.~\ref{sec:decomres}.}
\end{table} 

\citet{Watson2009} proposed to define crystalline indices to characterize the crystalline dust features, where the calculation of the indices is not dependent on any model assumption. Following their approach, we first removed numerous molecular lines in the spectra that are not the interest of this work. The line flux to continuum ratios are particularly high in some cases (Xie et al. 2026, submitted), given the excellent spectral resolution of MIRI ($R\,{\sim}\,2500$), see Figure~\ref{fig:obs}. To this end, we applied a median filtering method to obtain the overall shape of the continuum by setting the filter width to 95 wavelength channels, and only kept data points that deviate from the overall shape of the continuum by less than 1\%. This approach preserves dust features in the raw spectrum while effectively eliminating the influence from molecular lines. \citet{Pontoppidan2024} developed the \texttt{ctool} program to identify the underlying continuum through an iterative algorithm, see Sect.~4.1 of their work for details. We compared our results with the continuum returned from \texttt{ctool}, and found that the differences are negligible. Once the continuum is identified, we fit a third-order polynomial to the spectra in the wavelength range of $5.61\,{-}\,7.94\,\mu{\rm m}$, $13.02\,{-}\,13.50\,\mu{\rm m}$ and $14.32\,{-}\,14.83\,\mu{\rm m}$, within which silicate emission features are minor. As a next step, we normalized the spectrum to the fitted continuum via $F_{\nu,\rm{norm}}\,{=}\,(F_{\nu,\rm{obs}}\,{-}\,F_{\nu,\rm{cont}})/F_{\nu,\rm{cont}}$. Then, we defined the equivalent width of the $10\,\mu{\rm m}$ feature as $W_{10}\,{=}\,\int F_{\nu,\rm{norm}} d\nu$, where the integration is performed over a frequency ($\nu$) range equivalent to [$8\,\mu{\rm m}$, $12\,\mu{\rm m}$]. A same process was applied to LkCa\,15 and UY\,Aur, both of which show interstellar-like silicate features. The average of their normalized spectra represents the reference spectrum ($F_{\nu,\rm{norm},0}$) produced by small amorphous silicates. 

\begin{table}[!t]
\caption{Kendall’s $\tau$ coefficients and the corresponding probability $P$ values derived from correlation tests for parameter pairs shown in Figure~\ref{fig:index} and \ref{fig:correlation}.}
\centering
\linespread{1.3}\selectfont
\label{tab:kendall}
\begin{tabular}{lcc}
 \hline
\multirow{2}{*}{Parameter pair}  & $\tau$, $P$  &  $\tau$, $P$  \\
                           &  (Taurus, $N\,{=}\,31$)  & (USco, $N\,{=}\,13$)  \\
\hline
$P_{10}\,-\,O_{10}$  & 0.24, 0.06               &  $-$0.03, 0.9  \\
$W_{10}\,-\,P_{10}$  & $\mathbf{-0.40}$,  $\mathbf{<0.01}$      &  0.03, 0.9      \\
$W_{10}\,-\,O_{10}$  & $\mathbf{-0.42}$,  $\mathbf{<0.01}$      &  $\mathbf{-0.59}$, $\mathbf{<0.01}$  \\
$\dot{M}_{\rm acc}\,-$\,Crystallinity   & 0.09, 0.48     &  $-$0.19, 0.36  \\
$\dot{M}_{\rm acc}\,-$\,Grain size      & 0.21, 0.10     &  $-$0.16, 0.42  \\
$F_{24}/F_{8}\,-$\,Crystallinity        & $\mathbf{-0.31}$, {\bf 0.01}  &     0.00, 1.00  \\
$F_{24}/F_{8}\,-$\,Grain size           & $-$0.15, 0.25  &  $-$0.38, 0.07  \\
\hline
\end{tabular}
\end{table} 

We calculated the indices for the prominent features of pyroxene and olivine centered at $\lambda_{p}\,{=}\,9.21\,\mu{\rm m}$ and $\lambda_{o}\,{=}\,11.08\,\mu{\rm m}$ as 
\begin{equation}
 X = \frac{\int_{\nu_{x}-\Delta \nu}^{\nu_{x}+\Delta \nu} F_{\nu,\rm{norm}}d \nu} {\int_{\nu_{R}-\Delta \nu}^{\nu_{R}+\Delta \nu} F_{\nu,\rm{norm}}d \nu} \frac{\int_{\nu_{R}-\Delta \nu}^{\nu_{R}+\Delta \nu} F_{\nu,\rm{norm,0}}d \nu} {\int_{\nu_{x}-\Delta \nu}^{\nu_{x}+\Delta \nu} F_{\nu,\rm{norm,0}}d \nu}, \,\,\,\, X\,{=}\,P,\,\,O.
\end{equation}
The comparison was conducted near the peak of the amorphous silicate feature $\lambda_{R}=9.94\,\mu{\rm m}$, with the integration performed over a frequency interval corresponding to a wavelength width of
$2\Delta \lambda\,{=}\,0.545\,\mu{\rm m}$. If the emission profile is indicative of amorphous silicates, the index approaches unity, while increasing indices above unity points to stronger crystalline signature.  

The calculated crystalline indices and the $10\,\mu{\rm m}$ equivalent widths are presented in Figure~\ref{fig:index}.  We notice two outliers in these plots, J16141107 and J16153456. As mentioned in Sect.~\ref{sec:obs}, four of the targets in USco are binaries. Nevertheless, we do not distinguish between binary and single systems in the remainder of this work, because previous studies have shown that dust properties inferred from infrared spectra are broadly similar in single and multiple systems \citep[e.g.,][]{Pascucci2008,Furlan2011,Manoj2011}. Moreover, we did not exclude the two outliers from the statistical tests. They have no impact on the results, see Sect.~\ref{sec:outliers} for details. The pyroxene indices ($P_{10}$) range from ${\sim}\,1$ to ${\sim}\,2.6$, with the median being 1.24 and 1.21 for the USco and Taurus sample, respectively. The olivine indices ($O_{10}$) lie between ${\sim}\,1$ to ${\sim}\,2.5$. The median value for the older USco targets is 1.53, which is slightly larger than that of the younger Taurus group, i.e., 1.27. As given in Table~\ref{tab:statistic}, the K-S p value (0.66) for the pyroxene index ($P_{10}$) is greater than the chosen significance level of 0.05, suggesting that the two samples may have the same distribution. A comparison of the olivine index ($O_{10}$) yields the same result, although a relatively small p-value of 0.07 was obtained. The equivalent widths of the $10\,\mu{\rm m}$ complex ($W_{10}$) for the USco sample are concentrated in the region between 0.5 and 2.5, except for J16153456, which has a much larger value of 8.8. The K-S test returned a p-value of 0.69, implying that the $W_{10}$ arrays for the Taurus and USco disks are drawn from the same distribution. 

\citet{Watson2009} found a strong correlation between $P_{10}$ and $O_{10}$ in a large sample of 84 Taurus disks. This was interpreted as evidence that the mechanisms responsible for the crystallization at early stages do not favor one mineral type over another. Moreover, the crystallization process likely does not involve significant transformation or incorporation of one mineral family into another. Otherwise, an anticorrelation would be expected. As shown in the upper panel of Figure~\ref{fig:index}, our smaller Taurus sample hints at a trend. However, the associated $P$-value is slightly above the 5\% significance level, so the evidence remains inconclusive. The old USco sample does not even show a hint of a trend between the two quantities (see Table~\ref{tab:kendall}). A larger sample would be helpful to test if the correlation found by \citet{Watson2009} is erased at later stages.

The $10\,\mu{\rm m}$ feature mainly arises from small dust grains in the optically thin surface layer of the inner disk. The equivalent width ($W_{10}$) is a quantitative measure of the overall strength of the silicate feature, and tells how prominent the silicate emission is compared to the disk continuum underneath. The middle and bottom panels of Figure~\ref{fig:index} indicate that there is an anticorrelation between crystalline indices ($P_{10}$ and $O_{10}$) and $W_{10}$ for Taurus disks. Such a trend has been reported before, but in variant forms, where the feature strength and feature shape are quantified by the peak-over-contuum $10\,\mu{\rm m}$ region and the flux ratio between $11.3\,\mu{\rm m}$ and $9.8\,\mu{\rm m}$ measured in the normalized spectra respectively \citep[e.g.,][]{Apai2005,vanBoekel2005,Bouwman2008,Pascucci2009}. Comparing the opacities of dust grains of various sizes with observations, \citet{Kessler-Silacci2006} interpreted the correlation as a sign of grain growth. The strength of the emission bands with large equivalent widths is indicative of small and amorphous silicate grains, whereas larger grains and a higher degree of crystallinity on average produce small equivalent widths. \citet{Watson2009} noticed that $W_{10}$ is positively correlated with the infrared spectral slope that is a probe for dust settling. From comparisons with disk models in which the degree of sedimentation is parameterized as in \citet{DAlessio2006}, they found that the trend between crystalline index and silicate-feature equivalent width can also be explained by small grains (without grain growth) in disks with a range of dust settling to the midplane. Nevertheless, detailed modeling of the infrared spectra of a large sample of disks frequently show grain growth to a few microns \citep[e.g.,][]{Bouwman2008,Sargent2009,Olofsson2010}. Therefore, a combination of grain growth and settling may be a more plausible explanation for the observed correlations between the crystalline index and the $10\,\mu{\rm m}$ equivalent width. It should be noted that unlike the Taurus sample, the pyroxene index $P_{10}$ is uncorrelated with $W_{10}$ for USco disks. 

\section{Spectral decomposition}
\label{sec:analysis}

Crystallization produces complexes of small, sharp peaks superimposed on the relatively broad pristine dust features, which causes the entire feature to appear wider. Grain growth causes the silicate features to become wider and flatter as well. Although the crystalline indices and equivalent width are very useful and can be conveniently computed for a large sample of objects, it is difficult to disentangle the coupled effects of grain growth and crystallinity on the appearance of infrared spectra. In order to analyze the $10\,\mu{\rm m}$ silicate feature in a more quantitative way, we performed a detailed spectral decomposition for the Taurus and USco disks. This approach has been widely used in previous works \citep[e.g.,][]{vanBoekel2005,Honda2006,Bouwman2008,Sargent2009,Jang2024b}, and involves the establishment of model spectra and fitting to the data. The mass absorption coefficients of different dust components and sizes are incorporated into the model. An essential assumption is that the emergent spectrum can be reproduced by adding up the emission from each of the individual constituents.

\subsection{Dust model}
\label{sec:dustmodel}

\begin{figure*}
\centering
\includegraphics[width=0.9\textwidth, angle=0]{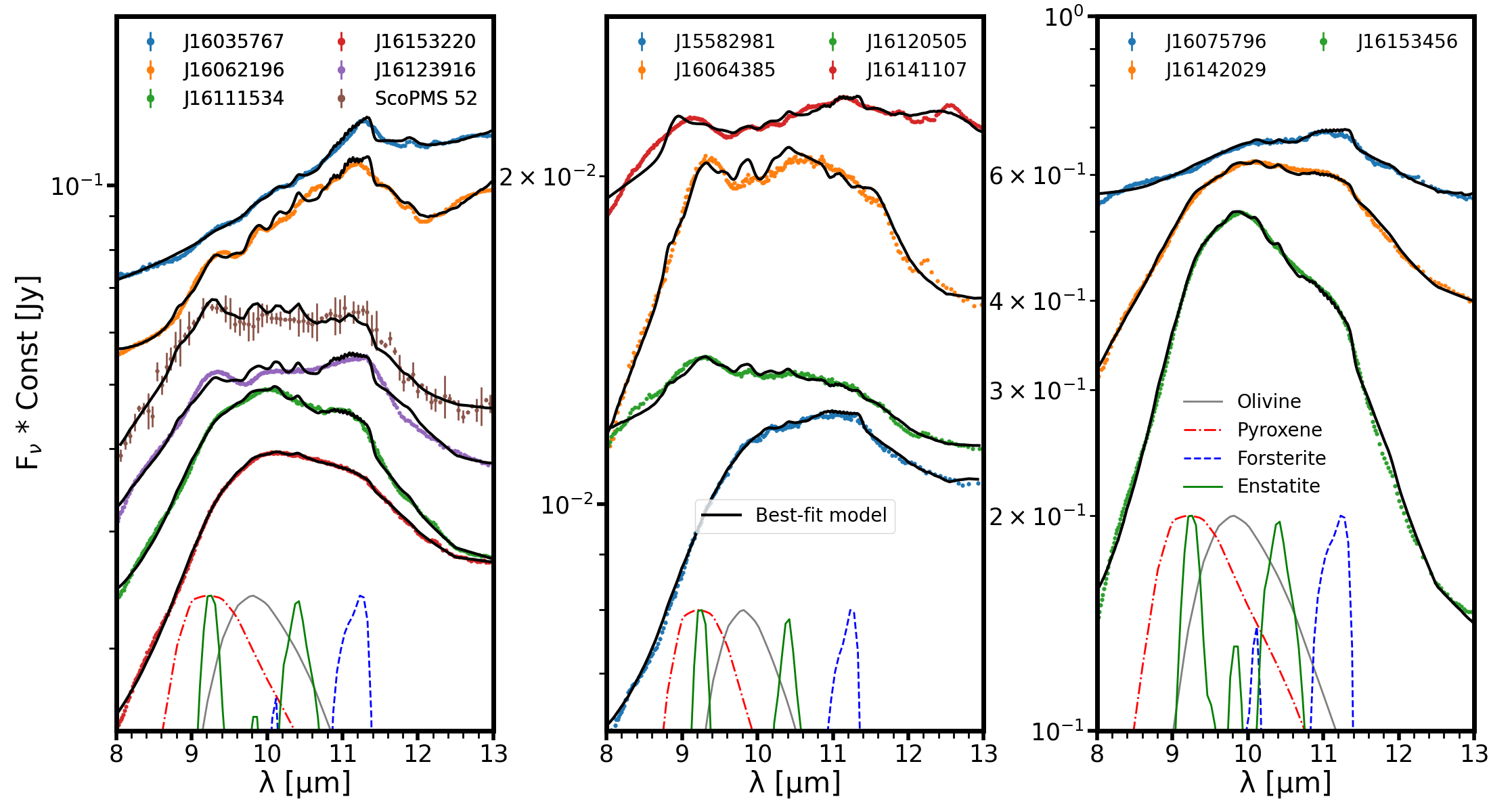}
\caption{Results of spectral decomposition for USco objects. The best-fit model is indicated with black curves, while colored dots refer to observations. The spectra are scaled for a better representation. In each panel, we display the normalized mass absorption coefficients for dust grains with a size of $0.1\,\mu{\rm m}$. The disks are arranged from top to bottom in order of decreasing dust crystallinity ($C$).}
\label{fig:uscofit}
\end{figure*}

\begin{figure*}
\centering
\includegraphics[width=0.9\textwidth, angle=0]{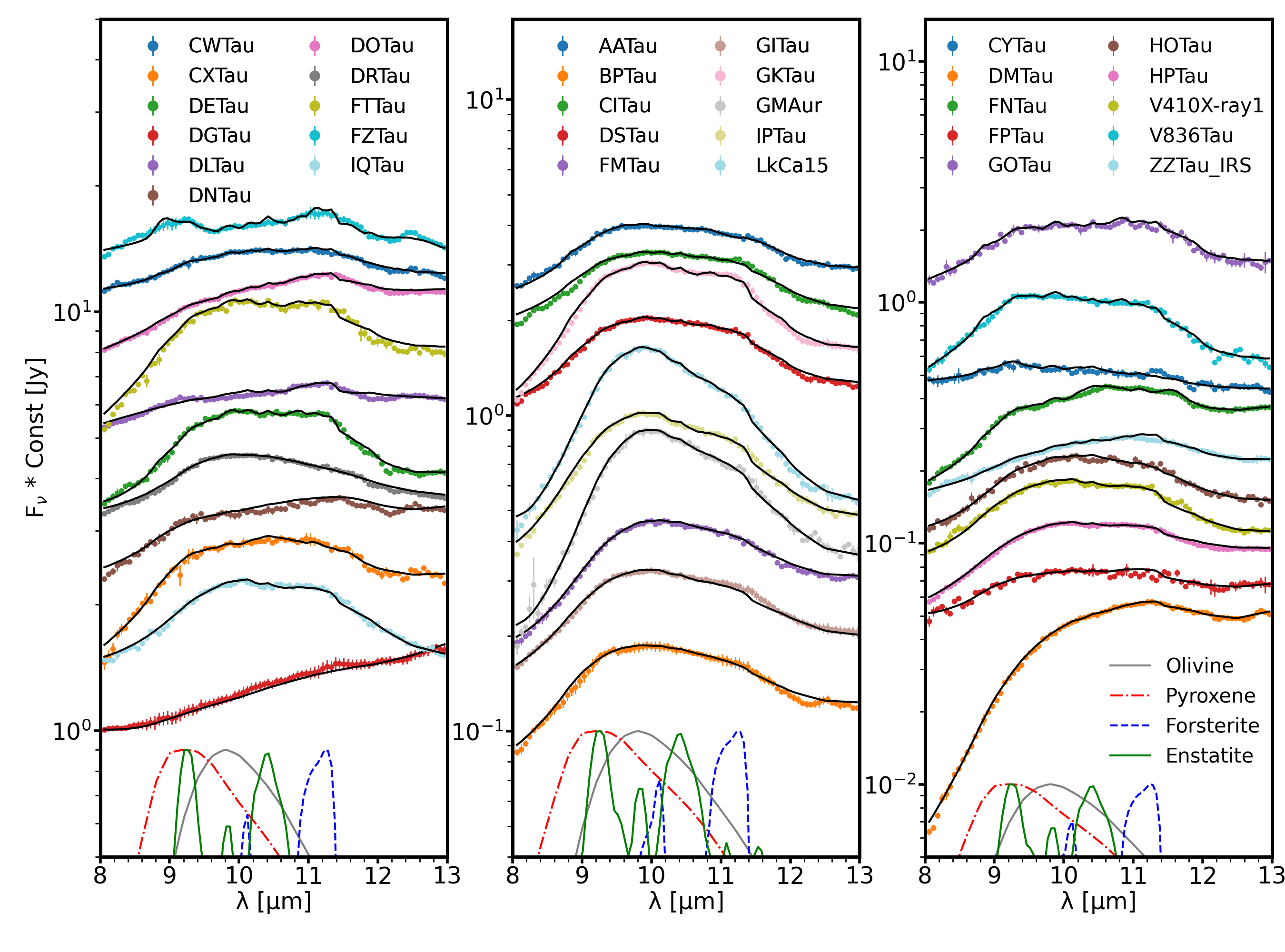}
\caption{Same as in Figure~\ref{fig:uscofit}, but for Taurus disks.}
\label{fig:taurusfit}
\end{figure*}

For the dust model, we selected four species: amorphous silicate with olivine stoichiometry \citep[${\rm MgFeSiO_{4}}$,][]{Dorschner1995}, amorphous silicate with pyroxene stoichiometry \citep[${\rm MgFeSi_{2}O_{6}}$,][]{Dorschner1995}, crystalline forsterite \citep[${\rm Mg_{2}SiO_{4}}$,][]{Servoin1973}, and crystalline enstatite \citep[${\rm MgSiO_{3}}$,][]{Jager1998}. These components have been commonly identified in infrared spectra of protoplanetary disks \citep[e.g.,][]{Bouwman2008,Sargent2009,Juhasz2010,Liu2025}. We assumed amorphous silicates to be homogeneous spheres and calculated the mass absorption coefficient using the Mie theory. For crystalline silicates, since their observed bands cannot be simply reproduced by homogeneous spherical particles \citep{2001A&A...375..950B}, we simulated their shape using a distribution of hollow spheres \citep{Min2005}. The irregularity parameter was set to $f_{\rm max}\,{=}\,0.9$. 

Dust grains of different sizes have different feature strengths and central wavelengths of their resonant bands, which in turn contribute differently to the spectrum. We considered three sizes for each type of dust, 0.1\,$\mu {\rm m}$, 2.0\,$\mu {\rm m}$ and 5.0\,$\mu {\rm m}$, and calculated the dust opacity using the \texttt{OpTool}\footnote{\url{https://github.com/cdominik/optool}} \citep{Dominik2021}. For both amorphous and crystalline silicates, the emission features gradually weaken with increasing particle size, suggesting that larger particles have lower influence on the spectrum. A gallery of the dust opacities and their variation with grain size can be found in Figure~3 of \citet{Bouwman2008}. 

\subsection{Spectral model and fitting approach}
The spectral model was built upon the Two-Temperature model introduced by \citet{Bouwman2008}. Besides the underlying continuum from the optically thick interior of the disk, we also considered the contribution from an inner rim. The model spectrum is parameterized via
\begin{equation}\label{eq1}
F_{\nu}= B_{\nu}(T_{\rm rim})C_{r} + B_{\nu}(T_{\rm cont})C_{c} + B_{\nu}(T_{\rm dust})(\sum_{i=1}^{N}\sum_{j=1}^{M}C_{i,j}\kappa_{\nu}^{i,j}),
\end{equation}
where $B_{\nu}(T_{\rm rim})$, $B_{\nu}(T_{\rm cont})$ and $B_{\nu}(T_{\rm dust})$ refer to the Planck function at the inner rim temperature $T_{\rm rim}$, continuum temperature $T_{\rm cont}$ and dust temperature $T_{\rm dust}$, respectively. In general, dust grains with different compositions and sizes can attain different equilibrium temperatures \citep{Krugel2008}. Nevertheless, we adopted the simplified assumption that all dust species share a single temperature, because introducing grain size- and composition-dependent dust temperatures would significantly increase the number of free parameters and computational costs, without necessarily improving the robustness of the conclusions within the scope of this work. The mass absorption coefficients for the dust of $j$ type and $i$ size are denoted as $\kappa_{\nu}^{i,j}$. The scaling factor of the rim emission is given by $C_{r}$, while $C_{c}$ and $C_{i,j}$ are weighting factors for the continuum and each of the dust components. Normalizing $C_{i,j}$ yields the mass fraction for each type of dust   
\begin{equation}\label{eq2}
m_{i,j}=\frac{C_{i,j}}{\sum_{j=1}^{N}\sum_{i=1}^{M}C_{i,j}}.
\end{equation}

Before performing the fitting, we subtracted the stellar contribution from the observed spectrum. The stellar spectrum is simply represented by a Planck function. For Taurus disks, we used the stellar properties reported by \citet{Andrews2013}, rescaling the stellar luminosities according to the Gaia distance, see Table~\ref{tab:target}. The stellar parameters for USco disks were taken from \citet{Fang2023}. We did not subtract the stellar emission for the four binaries in USco, because reliable estimates of the individual stellar contributions are not yet available. This simplification may affect the derived temperatures to some extent. However, it is not expected to significantly influence the inferred mass fractions of the different grain species. We examined the extinction values for all targets \citep{Andrews2013, Fang2023} and found that most exhibit low extinction, with $A_{\rm V}\,{\lesssim}\,1\,{\rm mag}$. Such low levels of extinction are expected to have only a minor effect on the $10\,\mu{\rm m}$ silicate feature, as extinction decreases significantly toward mid-infrared wavelengths. For example, an optical extinction of $A_{V}\,{=}\,1\,{\rm mag}$ corresponds to only ${\sim}\,0.01\,{\rm mag}$ at $\lambda\,{=}\,10\,\mu{\rm m}$ according to the extinction law of \citet{Cardelli1989} with $R_{V}\,{=}\,3.1$. Therefore, foreground extinction is neglected. For the parameter study, we utilized \texttt{UltraNest}\footnote{\url{{https://johannesbuchner.github.io/UltraNest/}}}, which is a Python package for nested sampling \citep[]{Skilling2004,Buchner2023}, and designed for efficient Bayesian inference \citep{2021Buchner}. It supports arbitrary user-defined vectorized likelihood functions, enhances uncertainty estimation, and retains additional live points during parallel execution, thereby reducing the computational cost of model evaluations. We selected the 8\,–\,13\,$\mu {\rm m}$ portion of the spectrum for fitting. This is because dust features are prominent in these domains, and therefore dust properties can be well constrained. 

\begin{figure*}
\centering
\includegraphics[width=\textwidth, angle=0]{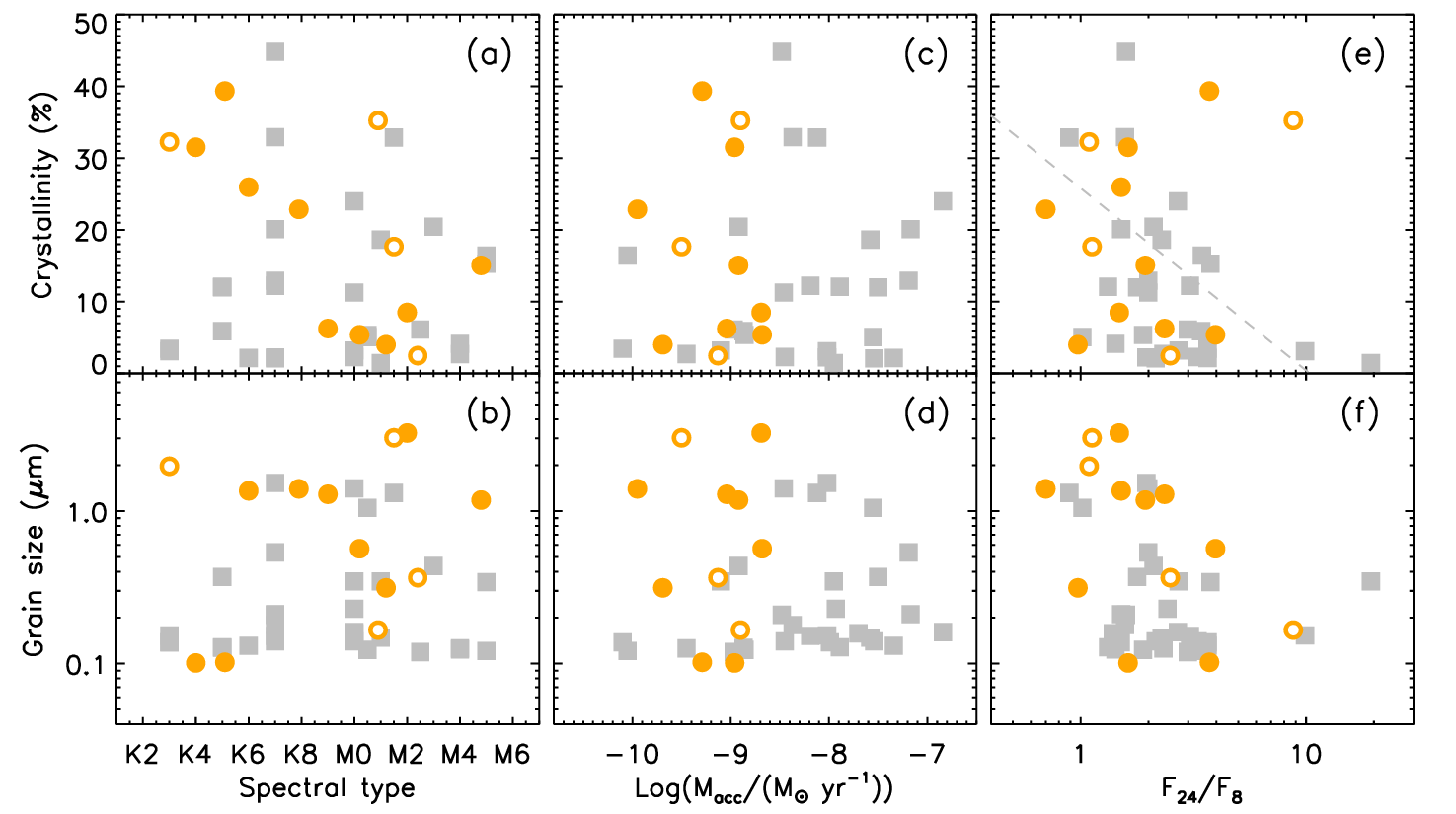}
\caption{Average sizes of amorphous grains (bottom panels) and crystallinity (upper panels) plotted versus the spectral type of the central stars (left column), accretion rate (middle column), and the flux ratio of $F_{24}/F_{8}$ as a measure of disk geometry (right column). As in Figure~\ref{fig:index}, gray squares represent Taurus disks, while orange dots and circles refer to USco disks around single stars and binaries, respectively. Table~\ref{tab:statistic} summarizes the results of K-S tests that compare the distribution of crystallinity and grain size between the Taurus and USco sample. The Kendall’s $\tau$ correlation coefficients and the corresponding $P$-values for each of the parameter pairs are provided in Table~\ref{tab:kendall}. The dashed line indicates a statistically significant trend (Kendall’s $P\,{<}\,0.05$).}
\label{fig:correlation}
\end{figure*}

\subsection{Results and discussion}
\label{sec:decomres}

The fitting procedure took about one hour for each object. The time cost for convergence mainly depends on the number of data points engaged in the fitting. A comparison between model and observation for USco disks is presented in Figure~\ref{fig:uscofit}, whereas Figure~\ref{fig:taurusfit} shows the result for Taurus disks. As can be seen, the models capture well the observed characteristics of the $10\,\mu{\rm m}$ feature. Table~\ref{tab:uscores} and \ref{tab:taurusres} summarize the inferred crystallinity and mass-averaged grain size (radius). The average size was derived using all of the considered dust species ($\langle a \rangle$) and only amorphous silicates ($\langle a_{\rm am} \rangle$), separately. The best-fit parameter sets together with their confidence intervals can be found in Table~\ref{tab:uscomfrac} and \ref{tab:taurusmfrac}. 

We performed a K-S test to check whether the distribution of average size of amorphous silicates and crystallinity differs between the young and old groups. The returned p-value (0.016) for the average size is smaller than the chosen significance level of 0.05, implying that the two samples are drawn from different distributions. The median average size of USco disks is $1.18\,\mu{\rm m}$, roughly six times larger than the median average size of $0.16\,\mu{\rm m}$ for Taurus disks, see Table~\ref{tab:statistic}. This is consistent with theories of dust evolution in protoplanetary disks, where grain growth proceeds in the inner disk on time scales shorter than a few Myrs \citep[e.g.,][]{Weidenschilling1997,Birnstiel2024}. The median crystallinity of the USco sample is ${\sim}\,18\%$, while it is about $12\%$ for the Taurus group. The two samples appear indistinguishable in terms of crystallinity as suggested by a large p-value of 0.59 from the K-S test. The lack of correlation between crystalline silicates and stellar age was also noted by \citet{Sicilia-Aguilar2007} and \citet{Watson2009b}. \citet{Kessler-Silacci2006} calculated the peak-over-continuum $10\,\mu{\rm m}$ region and normalized flux ratio $F_{11.3}/F_{9.8}$ for a sample of 47 disks observed as part of the Core to Planet-Forming Disk Spitzer legacy program \citep{Evans2003}. Their sample is 0.5$-$6\,Myr old, and also show no strong relation between the strength-shape trend and the stellar age. Therefore, our work extends the literature finding to more evolved systems with ages of 5$-$10\,Myr. 

In order to complete our understanding of the differences in silicate emission due to stellar properties and disk structure, we compared the average grain size and crystallinity with spectral type, accretion rate and flux ratio $F_{24}/F_{8}$ between $24\,\mu{\rm m}$ and $8\,\mu{\rm m}$. The results are presented in Figure~\ref{fig:correlation}. As can be seen from panel (a), the crystallinity appears unrelated to the spectral type for Taurus disks with a relatively narrow range of spectral type, consistent with previous findings \citep{Watson2009,Olofsson2010}. However, the USco sample seems to follow a decreasing crystallinity around later type stars. This may be explained in terms of the increasing disk temperature towards earlier type stars, leading to a larger region in which the disk material is sufficiently hot for thermal annealing (and chemical equilibrium processes) to occur \citep[e.g.,][]{Fabian2000,Jager2003}. The crystallinity is obtained from fitting the $10\,\mu{\rm m}$ feature that probe the inner and warm disk region. Compared with the USco sample, the accretion rate of Taurus disks studied in this work is systematically higher, see panel (c) of Figure~\ref{fig:correlation}. Therefore, disk wind and turbulent mixing \citep[e.g.,][]{Abraham2009,Pascucci2023pp7}, which are believed to radially transport crystals to the outer disk and therefore weaken the crystallinity-stellar type relation, works more efficiently. Consequently, the trend in the crystallinity-stellar type diagram is not preserved for the Taurus sample. MIRI observations towards more USco disks with a broader range of spectral type are necessary for confirming our speculation. 

There is a hint of a trend between the average grain size and spectral type in panel (b) of Figure~\ref{fig:correlation}. Larger grains may be more prevalent around later-type stars, although the evidence remains limited. The tentative nature of this relationship is likely due to the relatively narrow range of spectral types explored in this study. A clear trend is identified by previous works when the coverage of spectral type is large to include A/B stars and/or brown dwarfs \citep[e.g.,][]{Apai2005,Kessler-Silacci2006,Pascucci2009}. Generally speaking, disks around later type (lower mass) stars are cooler than those of their higher mass counterparts. Therefore, the disk region probed by the $10\,\mu{\rm m}$ feature for later type stars is closer to the stellar host, where the density is higher. Grain growth proceeds faster at higher density \citep{Birnstiel2024}. Moreover, dust settling that removes large dust grains from disk surface layers decreases at higher density \citep{Dullemond2004a}. Thus, the combination of faster grain growth and slower dust settling at smaller radii could explain the increasing grain size towards later type stars. 

The crystallinity and average grain size show no clear dependence on the accretion rate. The upper boundary of grain sizes in the Taurus disks may show a tentative increase with accretion rate up to $\dot{M}_{\rm acc}\,{\sim}\,10^{-8}\,M_{\odot}\,{\rm yr}^{-1}$, followed by a possible decrease at higher accretion rates, although this pattern is not statistically significant given the small number of objects. In other words, the largest grains may preferentially occur in systems with moderate-to-high accretion rates, while disks with the highest accretion rates may tend to exhibit somewhat smaller grain sizes in their surface layers. A similar trend is seen in Figure~14 of \citet{Sicilia-Aguilar2007}, who analyzed IRS spectra of a sample of disks in the ${\sim}\,4\,{\rm Myr}$-old Tr 37 cluster. Disks with higher accretion rates are generally expected to be more turbulent \citep[e.g.,][]{Hartmann1998,Armitage2011}, allowing large grains to be stirred up from the disk interior to surface layers. However, strong turbulence in highly accreting disks may also enhance grain–grain collisions, leading to more efficient fragmentation and thus a reduction in the characteristic grain size. Given the limited number of objects in this regime, this apparent turnover should be regarded as suggestive rather than statistically significant. Nevertheless, a similar trend in the upper boundary of grain sizes is not evident in the more evolved USco disks.

One particularly important question in the field of planet formation is how dust processing is related to the evolution of disk structures. This topic has been widely explored in the literature. Infrared spectral indices or flux ratios between two wavelengths where dust features are weak have been commonly used to probe disk flaring (or dust settling). Radiative transfer simulations with a sophisticated treatment of dust sedimentation show steeply decreasing emission toward longer wavelengths and lower flux ratios in more settled disks \citep{Dullemond2004a,DAlessio2006}. By analyzing the IRS spectra of a sample of Taurus disks, \citet{Watson2009} showed that the spectral index measured between $13\,\mu{\rm m}$ and $31\,\mu{\rm m}$ ($n_{13-31}$) correlates with the crystalline olivine index $O_{10}$ and the $10\,\mu{\rm m}$ equivalent width $W_{10}$. The dependence of $n_{13-31}$ on $W_{10}$ has also been found in large samples of disks in the Chamaeleon I \citep{Manoj2011} and Ophiuchus star-forming regions \citep{McClure2010}. Regardless of whether grain growth or crystallization is the dominant mechanism shaping the spectral features, the presence of these trends suggests that dust processing proceeds in tandem with the evolution of the disk. Through decomposing the IRS spectra of 65 Taurus disks, \citet{Sargent2009} found higher crystallinity in more settled disks. This behavior is reproduced by our analysis (see panel (e) of Figure~\ref{fig:correlation} and Table~\ref{tab:kendall}). The $F_{24}/F_{8}$ flux ratio serves as a proxy for dust settling, with lower values pointing to more settled disks. A similar trend between crystallinity and dust settling has also been reported in a sample of six disks around brown dwarfs \citep{Apai2005}. However, the correlation between crystallinity and dust settling is not detected in USco disks according to the Kendall rank correlation coefficient and significance level (Table~\ref{tab:kendall}). As illustrated in panel (f) of Figure~\ref{fig:correlation}, large grains appear to be predominantly present in relatively flat disks with low $F_{24}/F_{8}$ ratios, whereas smaller grains are found in both flared and flat disks. This behavior is consistent with findings from previous works \citep[e.g.,][]{Bouwman2008,Juhasz2010,Olofsson2010}, suggesting that grain growth can occur across a wide range of disk geometries. However, the number of USco disks in our sample is relatively small, which limits the statistical significance of the observed distribution. With such a limited sample, any underlying correlation may remain undetected. Expanding the sample of disks in evolved star-forming regions will therefore be essential for more robustly assessing the relationship between grain growth and dust scale height at later stages of disk evolution.

\section{Long wavelength portion of the MIRI spectrum}
\label{sec:longwav}

\begin{figure}
\centering
\includegraphics[width=0.5\textwidth, angle=0]{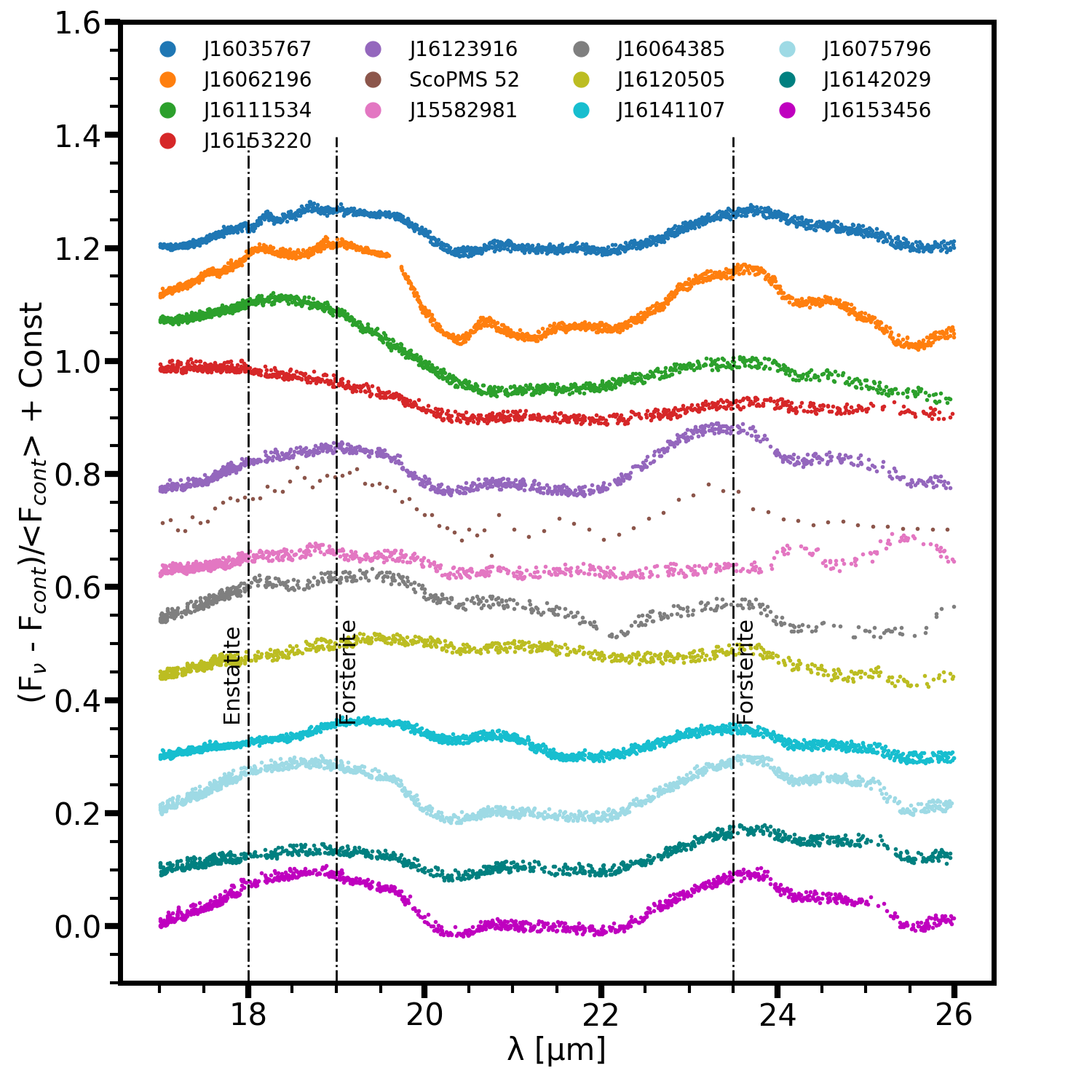}
\caption{Long wavelength portion of the MIRI spectra of the USco disks. The locations of promiment crystalline features are indicated with vertical lines. The spectra are shifted for a better visualization.}
\label{fig:longwav}
\end{figure}

We have fitted the IRS spectra of Taurus disks and MIRI spectra of USco targets in the vicinity of the $10\,\mu{\rm m}$ feature for comparisons with literature results which mostly focus on this feature. The long wavelength portion of the spectrum probes disk regions that are further out from the host central star. Figure~\ref{fig:longwav} shows the normalized spectra of USco disks, $F_{\nu,\rm{norm}}\,{=}\,(F_{\nu,\rm{obs}}\,{-}\,F_{\nu,\rm{cont}})/F_{\nu,\rm{cont}}$, in the wavelength range from $17\,\mu{\rm m}$ to $26\,\mu{\rm m}$. The locations of prominent features from crystalline forsterite and enstatite are labeled. The contrast of emission features at long wavelengths is obviously lower than that of the $10\,\mu{\rm m}$ feature. Therefore, we left the comparison of dust mineralogy revealed by long wavelength spectra to future works when a homogeneous MIRI dataset are available for both the Taurus and USco disks. 

In the long wavelength regime, the emission features from crystalline forsterite and enstatite are relatively broad and mixed with each other compared to that of the $10\,\mu{\rm m}$ feature, leading to different polynomials fitting the same spectrum equally well. This means that a calculation of crystalline index for each type of dust as conducted in Sect.~\ref{sec:cryidx} is not practical. Moreover, the relative strengths seen from the plot should also be treated with care, because they are highly affected by the fitted continuum. Therefore, the figure mainly exhibits which type of dust is detected for each object. As can be seen, most of the objects show crystalline forsterite features at ${\sim}\,19\,\mu{\rm m}$ and ${\sim}\,23\,\mu{\rm m}$ when they also exhibit crystalline characteristics at ${\sim}\,10\,\mu{\rm m}$, consistent with findings of previous works \citep[e.g.,][]{McClure2010,Furlan2011}. The fact that crystalline silicate features are detected at both short and long wavelengths implies that crystallization is either an early, disk-wide process or that efficient radial transport redistributes crystals from the inner disk to larger radii.  

\section{Summary}
\label{sec:summary}

In this work, we present and analyze new JWST/MIRI spectra of 11 protoplanetary disks in the 5–10 Myr-old USco association. The sample is supplemented with one disk from the JWST MINDS survey and one additional disk observed as part of the FEPS Spitzer legacy program. To investigate dust properties as a function of evolutionary stage, we construct a comparison sample of 31 disks in the 1–3 Myr-old Taurus Molecular Cloud with similar stellar spectral types and analyze their Spitzer/IRS spectra. We characterize dust properties using crystalline feature indices and further constrain dust crystallinity and grain size through spectral decomposition. A systematic comparison of dust properties between the USco and Taurus samples yields the following key results.

(1) No significant variation in crystallinity with cloud age. We find that the crystalline olivine index ($O_{10}$), crystalline pyroxene index ($P_{10}$) and the inferred crystallinity ($C$) are comparable between the Taurus and Usco disks, see Table~\ref{tab:statistic}. Specifically, the median crystallinity of Taurus disks is 12\%, while it is 17\% for Usco disks. This result suggests that dust crystallization is largely established at early evolutionary stages.

(2) Evidence of grain growth with cloud age. The median grain size of the USco disks is ${\sim}\,1.2\,\mu{\rm m}$, which is approximately six times larger than that of the Taurus disks ${\sim}\,0.2\,\mu{\rm m}$, see Table~\ref{tab:statistic}. This indicates continued growth of dust grains in the inner disk over time, consistent with theoretical models of dust evolution. More robust evidence for grain growth will require a homogeneous JWST/MIRI dataset for both young and evolved disks with a larger sample size.

(3) Coupled evolution of dust processing and dust scale height. For the Taurus disks, we find that crystallinity increases toward more settled disks, consistent with previous studies suggesting that dust processing proceeds alongside disk evolution. In contrast, this correlation is not detected for USco disks. Instead, USco disks show a possible tendency for larger grains to occur in more settled disks, although the limited sample size prevents a statistically robust conclusion. These results suggest that the relationship between dust processing and dust scale height may evolve with disk age.

Future JWST/MIRI observations will enable detailed mineralogical studies of a much larger number of disks in evolved star-forming regions such as USco. Expanding the sample with high-quality mid-infrared spectra will allow more robust statistical tests of the connections among stellar properties, dust processing and disk structural evolution, thereby providing stronger constraints on the timescales and mechanisms of dust evolution in protoplanetary disks.

\begin{acknowledgments}
We thank the anonymous referee for the constructive comments that highly improved the manuscript. Y.L. acknowledges financial supports by the Natural Science Foundation of Sichuan Province of China (grant no. 2025ZNSFSC0060), the Fundamental Research Funds for the Central Universities (grant no. 2682025CX028), the International Partnership Program of Chinese Academy of Sciences (grant no. 019GJHZ2023016FN), and the Natural Science Foundation of China (grant no. 11973090). This work is based on observations made with the NASA/ESA/CSA James Webb Space Telescope. These observations are associated with JWST GO Cycle 2 program ID 2970 (PI: I. Pascucci). Support for C.X. and I.P. through this program was provided by NASA through a grant from the Space Telescope Science Institute, which is operated by the Association of Universities for Research in Astronomy, Inc., under NASA contract NAS 5-03127. C.X. and I.P. also acknowledge partial support from the National Aeronautics and Space Administration under agreement No. 80NSSC21K0593 for the program ``Alien Earths".  R.B. and T.M. were supported by the Royal Society, award numbers URF\textbackslash R1\textbackslash 211799 and RF\textbackslash ERE\textbackslash 231082. The JWST data presented in this article were obtained from the Mikulski Archive for Space Telescopes (MAST) at the Space Telescope Science Institute. The specific observations analyzed can be accessed via \dataset[doi: 10.17909/mjkr-rt82]{https://doi.org/10.17909/mjkr-rt82}. This work is based on observations made with the Spitzer Space Telescope, which was operated by the Jet Propulsion Laboratory, California Institute of Technology under a contract with NASA. We thank Eshan Raul for insightful discussions. 
\end{acknowledgments}

%\begin{contribution}

%All authors contributed equally to the Terra Mater collaboration.

%% But authors are expected to provide more specific details, e.g. 
%%
%%SC was responsible for writing and submitting the manuscript.
%%WWM came up with the initial research concept and edited the manuscript.
%%OTS obtained the funding and edited the manuscript.
%%EBF provided the formal analysis and validation. He also edited the manuscript.
%%GEH Supervised the undergraduates, wrote the software and administers the project github and Zenodo repositories.
%%
%% Authors can use the Contributor Role Taxonomy (CRediT) at
%% for ideas on how write a good statement tailored to their needs.

%\end{contribution}

\facilities{JWST(MIRI), Spitzer(IRS)}

\software{Ultranest \citep{2021Buchner}
          }


%% For this sample we use BibTeX plus aasjournalv7.bst to generate the

\bibliography{usco}{}
\bibliographystyle{aasjournalv7}



\appendix

\section{More statistical tests for the USco sample}
\label{sec:outliers}

In the main sections of this work, all 13 USco disks were included in the statistical analysis. As shown in Figure~\ref{fig:index}, J16141107 exhibits a large pyroxene index ($P_{10}\,{=}\,2.6$), while J16153456 shows a particularly large equivalent width ($W_{10}\,{=}\,8.9$). Both objects can therefore be regarded as outliers. We therefore repeated all statistical tests after excluding these two targets, with the results presented in Tables~\ref{tab:kscom} and \ref{tab:kendalcom}.

By comparing Table~\ref{tab:kscom} with Table~\ref{tab:statistic} and Table~\ref{tab:kendalcom} with Table~\ref{tab:kendall}, we found that the statistical results for the full and reduced samples do not differ significantly.

\begin{table}[!h]
\caption{Median level of dust properties and the K$-$S test p values.}
\centering
\label{tab:kscom}
\begin{tabular}{ccccccc}
 \hline
Sample & $N$ & $O_{10}$ & $P_{10}$ & $W_{10}$ & $C$ & $\langle a_{\rm am} \rangle$  \\
\hline
USco    & 11  & 1.53 & 1.24 & 1.51 & 17.69\% & $1.18\,\mu{\rm m}$ \\
Taurus  & 31 & 1.27 & 1.21 & 1.91 & 12.00\% & $0.16\,\mu{\rm m}$  \\
\hline
K$-$S p  & $-$ & 0.09 & 0.82 & 0.74 & 0.49 & 0.05 \\
\hline
\end{tabular}
\end{table} 

\begin{table}[!h]
\caption{Kendall’s $\tau$ coefficients and the corresponding probability $P$ values derived from correlation tests for each parameter pair.}
\centering
\linespread{1.3}\selectfont
\label{tab:kendalcom}
\begin{tabular}{lcc}
 \hline
\multirow{2}{*}{Parameter pair}  & $\tau$, $P$  &  $\tau$, $P$  \\
                                 &  (Taurus, $N=31$)    &  (USco, $N=11$)      \\
\hline
$P_{10}\,-\,O_{10}$  & 0.24, 0.06               &  $-$0.24, 0.31  \\
$W_{10}\,-\,P_{10}$  & $-$0.40,  $<0.01$        &  0.27, 0.24      \\
$W_{10}\,-\,O_{10}$  & $-$0.42,  $<0.01$        &  $-$0.45, 0.05  \\
$\dot{M}_{\rm acc}\,-$\,Crystallinity   & 0.09, 0.48     &  $-$0.02, 0.94  \\
$\dot{M}_{\rm acc}\,-$\,Grain size      & 0.21, 0.10     &  $-$0.09, 0.70  \\
$F_{24}/F_{8}\,-$\,Crystallinity        & $-$0.31, 0.01  &  0.16, 0.48     \\
$F_{24}/F_{8}\,-$\,Grain size           & $-$0.15, 0.25  &  $-$0.42, 0.07  \\
\hline
\end{tabular}
\end{table} 

\section{More information about the best-fit model parameters}
In this section, we present the best-fit parameter set for each object. The derived dust crystallinity and mass-averaged grain sizes are summarized in Tables~\ref{tab:uscores} and \ref{tab:taurusres}. Tables~\ref{tab:uscomfrac} and \ref{tab:taurusmfrac} list the rim temperature, continuum temperature, dust temperature, and the mass fractions of the different dust species.

\begin{table}[!h]
\caption{Derived dust crystallinity and mass-averaged grain sizes of USco disks.}
\centering
\linespread{1.5}\selectfont
\label{tab:uscores}
\begin{tabular}{cccc}
 \hline
\multirow{2}{*}{Object} & $C$  & $\langle a \rangle$ & $\langle a_{\rm am} \rangle$  \\
   &      $(\%)$     &    ($\mu{\rm m}$)   &   ($\mu{\rm m}$) \\
\hline\noalign{\smallskip}  
J15582981 &$\mathrm{15.06}_{-4.64}^{+4.13}$& $\mathrm{1.71}_{-0.72}^{+0.64}$& $\mathrm{1.18}_{-0.74}^{+0.65}$  \\ 
J16035767 &$\mathrm{39.35}_{-0.87}^{+0.86}$& $\mathrm{0.43}_{-0.02}^{+0.02}$& $\mathrm{0.10}_{-0.01}^{+0.00}$  \\ 
J16062196 &$\mathrm{35.24}_{-0.32}^{+0.31}$& $\mathrm{0.50}_{-0.02}^{+0.02}$& $\mathrm{0.17}_{-0.04}^{+0.04}$  \\ 
J16064385 &$\mathrm{22.87}_{-0.43}^{+0.44}$& $\mathrm{1.44}_{-0.04}^{+0.04}$& $\mathrm{1.40}_{-0.05}^{+0.06}$  \\ 
J16075796 &$\mathrm{31.52}_{-2.57}^{+1.40}$& $\mathrm{1.50}_{-0.15}^{+0.08}$& $\mathrm{0.10}_{-0.01}^{+0.00}$  \\ 
J16111534 &$\mathrm{4.00}_{-0.28}^{+0.14}$& $\mathrm{0.31}_{-0.03}^{+0.01}$& $\mathrm{0.31}_{-0.03}^{+0.01}$  \\ 
J16120505 &$\mathrm{17.69}_{-0.40}^{+0.40}$& $\mathrm{2.82}_{-0.16}^{+0.16}$& $\mathrm{3.03}_{-0.20}^{+0.19}$  \\ 
J16123916 &$\mathrm{8.50}_{-0.08}^{+0.08}$& $\mathrm{3.09}_{-0.06}^{+0.06}$& $\mathrm{3.25}_{-0.07}^{+0.07}$  \\ 
J16141107 &$\mathrm{32.26}_{-1.48}^{+0.67}$& $\mathrm{1.66}_{-0.12}^{+0.07}$& $\mathrm{1.97}_{-0.18}^{+0.10}$  \\ 
J16142029 &$\mathrm{6.26}_{-0.05}^{+0.05}$& $\mathrm{1.28}_{-0.01}^{+0.01}$& $\mathrm{1.29}_{-0.01}^{+0.01}$   \\ 
J16153220 &$\mathrm{2.48}_{-0.10}^{+0.08}$& $\mathrm{0.37}_{-0.02}^{+0.01}$& $\mathrm{0.37}_{-0.02}^{+0.01}$  \\ 
J16153456 &$\mathrm{5.39}_{-0.28}^{+0.30}$& $\mathrm{0.60}_{-0.05}^{+0.05}$& $\mathrm{0.57}_{-0.05}^{+0.05}$  \\ 
ScoPMS 52 &$\mathrm{25.95}_{-5.36}^{+5.37}$& $\mathrm{1.35}_{-0.43}^{+0.48}$& $\mathrm{1.36}_{-0.56}^{+0.63}$  \\ 
\hline
\end{tabular}
\end{table}

\begin{table}
\caption{Derived dust crystallinity and mass-averaged grain sizes of Taurus disks.}
\centering
\linespread{1.5}\selectfont
\label{tab:taurusres}
\begin{tabular}{cccc}
 \hline
\multirow{2}{*}{Object} & $C$  & $\langle a \rangle$ & $\langle a_{\rm am} \rangle$  \\
   &      $(\%)$     &    ($\mu{\rm m}$)   &   ($\mu{\rm m}$) \\
\hline\noalign{\smallskip}
AA Tau &$\mathrm{44.81}_{-31.94}^{+33.47}$& $\mathrm{2.19}_{-1.81}^{+1.95}$& $\mathrm{0.21}_{-0.16}^{+0.18}$  \\ 
BP Tau &$\mathrm{2.11}_{-1.85}^{+1.35}$& $\mathrm{0.17}_{-0.20}^{+0.14}$& $\mathrm{0.14}_{-0.17}^{+0.12}$  \\ 
CI Tau &$\mathrm{12.91}_{-15.66}^{+11.45}$& $\mathrm{0.79}_{-0.86}^{+0.71}$& $\mathrm{0.54}_{-0.48}^{+0.46}$  \\ 
CW Tau &$\mathrm{55.61}_{-18.65}^{+18.24}$& $\mathrm{2.45}_{-1.12}^{+1.11}$& $\mathrm{0.14}_{-0.07}^{+0.08}$  \\ 
CX Tau &$\mathrm{6.11}_{-4.06}^{+2.19}$& $\mathrm{0.15}_{-0.12}^{+0.07}$& $\mathrm{0.12}_{-0.09}^{+0.05}$  \\ 
CY Tau &$\mathrm{32.88}_{-5.10}^{+6.56}$& $\mathrm{1.51}_{-0.45}^{+0.59}$& $\mathrm{1.32}_{-0.60}^{+0.79}$  \\ 
DE Tau &$\mathrm{18.65}_{-8.08}^{+7.93}$& $\mathrm{0.44}_{-0.25}^{+0.25}$& $\mathrm{0.15}_{-0.10}^{+0.10}$  \\ 
DG Tau &$\mathrm{2.16}_{-2.60}^{+2.24}$& $\mathrm{0.17}_{-0.24}^{+0.14}$& $\mathrm{0.13}_{-0.20}^{+0.10}$  \\ 
DL Tau &$\mathrm{20.11}_{-27.45}^{+16.07}$& $\mathrm{0.22}_{-0.37}^{+0.28}$& $\mathrm{0.21}_{-0.45}^{+0.33}$  \\ 
DM Tau &$\mathrm{1.45}_{-0.84}^{+0.71}$& $\mathrm{0.36}_{-0.23}^{+0.16}$& $\mathrm{0.35}_{-0.23}^{+0.15}$  \\ 
DN Tau &$\mathrm{11.28}_{-11.53}^{+8.75}$& $\mathrm{1.66}_{-0.81}^{+0.76}$& $\mathrm{1.41}_{-0.61}^{+0.65}$  \\ 
DO Tau &$\mathrm{23.98}_{-17.97}^{+11.28}$& $\mathrm{0.48}_{-0.56}^{+0.38}$& $\mathrm{0.16}_{-0.21}^{+0.15}$  \\ 
DR Tau &$\mathrm{12.00}_{-17.23}^{+10.67}$& $\mathrm{0.73}_{-0.95}^{+0.75}$& $\mathrm{0.37}_{-0.42}^{+0.45}$  \\ 
DS Tau &$\mathrm{12.09}_{-5.48}^{+4.91}$& $\mathrm{0.32}_{-0.18}^{+0.17}$& $\mathrm{0.13}_{-0.08}^{+0.09}$  \\ 
FM Tau &$\mathrm{2.29}_{-1.26}^{+0.86}$& $\mathrm{0.16}_{-0.09}^{+0.09}$& $\mathrm{0.14}_{-0.08}^{+0.08}$  \\ 
FN Tau &$\mathrm{16.44}_{-7.21}^{+4.54}$& $\mathrm{0.29}_{-0.17}^{+0.10}$& $\mathrm{0.12}_{-0.07}^{+0.05}$  \\ 
FP Tau &$\mathrm{2.71}_{-1.82}^{+1.55}$& $\mathrm{0.15}_{-0.15}^{+0.13}$& $\mathrm{0.13}_{-0.14}^{+0.12}$  \\ 
FT Tau &$\mathrm{20.46}_{-16.12}^{+12.59}$& $\mathrm{0.84}_{-0.85}^{+0.63}$& $\mathrm{0.44}_{-0.52}^{+0.32}$  \\ 
FZ Tau &$\mathrm{57.37}_{-13.79}^{+9.15}$& $\mathrm{0.50}_{-0.22}^{+0.15}$& $\mathrm{0.16}_{-0.10}^{+0.09}$  \\ 
GI Tau &$\mathrm{2.25}_{-2.14}^{+1.84}$& $\mathrm{1.54}_{-3.60}^{+2.16}$& $\mathrm{1.53}_{-3.67}^{+2.21}$  \\ 
GK Tau &$\mathrm{12.22}_{-4.77}^{+4.33}$& $\mathrm{0.33}_{-0.17}^{+0.17}$& $\mathrm{0.15}_{-0.11}^{+0.11}$  \\ 
GM Aur &$\mathrm{3.10}_{-5.12}^{+2.47}$& $\mathrm{0.24}_{-0.30}^{+0.19}$& $\mathrm{0.15}_{-0.13}^{+0.13}$  \\ 
GO Tau &$\mathrm{54.15}_{-25.54}^{+24.77}$& $\mathrm{1.12}_{-0.68}^{+0.66}$& $\mathrm{0.23}_{-0.22}^{+0.23}$  \\ 
HO Tau &$\mathrm{5.36}_{-2.58}^{+1.78}$& $\mathrm{0.15}_{-0.09}^{+0.07}$& $\mathrm{0.12}_{-0.08}^{+0.06}$  \\ 
HP Tau &$\mathrm{3.44}_{-2.00}^{+1.70}$& $\mathrm{0.18}_{-0.11}^{+0.10}$& $\mathrm{0.14}_{-0.08}^{+0.07}$  \\ 
IP Tau &$\mathrm{3.21}_{-2.04}^{+1.70}$& $\mathrm{0.37}_{-0.55}^{+0.38}$& $\mathrm{0.35}_{-0.56}^{+0.39}$  \\ 
IQ Tau &$\mathrm{5.07}_{-8.28}^{+5.38}$& $\mathrm{1.14}_{-1.11}^{+0.98}$& $\mathrm{1.05}_{-1.07}^{+0.97}$  \\ 
LkCa 15 &$\mathrm{5.91}_{-1.97}^{+1.72}$& $\mathrm{0.21}_{-0.10}^{+0.10}$& $\mathrm{0.13}_{-0.07}^{+0.07}$  \\ 
V410 X-ray 1 &$\mathrm{4.13}_{-4.60}^{+1.86}$& $\mathrm{0.17}_{-0.14}^{+0.08}$& $\mathrm{0.12}_{-0.07}^{+0.05}$  \\ 
V836 Tau &$\mathrm{32.93}_{-12.55}^{+13.17}$& $\mathrm{0.72}_{-0.34}^{+0.37}$& $\mathrm{0.18}_{-0.11}^{+0.13}$  \\ 
ZZ Tau IRS &$\mathrm{15.29}_{-15.30}^{+5.56}$& $\mathrm{0.48}_{-0.98}^{+0.26}$& $\mathrm{0.34}_{-1.07}^{+0.26}$  \\ 
\hline
\end{tabular}
\end{table}

\begin{table*}[!h]
\caption{Retrieved temperature and mass fractions of each dust species for USco disks.}
\fontsize{40}{90}\selectfont
\label{tab:uscomfrac}
\rotatebox{0}{%
\resizebox{\textwidth}{!}{
\begin{tabular}{cccccccccccccccccccc}
\hline
 \multirow{2}{*}{Object}
&\multirow{2}{*}{$T_{\rm rim}$ (K)} 
&\multirow{2}{*}{$T_{\rm cont}$ (K)}  
&\multirow{2}{*}{$T_{\rm dust}$ (K)}  
&\multicolumn{3}{c}{Olivine (\%)}&
&\multicolumn{3}{c}{Pyroxene (\%)}&
&\multicolumn{3}{c}{Forsterite (\%)}&
&\multicolumn{3}{c}{Enstatite (\%)}\\
\cline{5-7}\cline{9-11}\cline{13-15}\cline{17-19}&&&
&0.1\,$\mu {\rm m}$  & 1.5\,$\mu {\rm m}$  & 6.0\,$\mu {\rm m}$&
&0.1\,$\mu {\rm m}$  & 1.5\,$\mu {\rm m}$  & 6.0\,$\mu {\rm m}$&
&0.1\,$\mu {\rm m}$  & 1.5\,$\mu {\rm m}$  & 6.0\,$\mu {\rm m}$&
&0.1\,$\mu {\rm m}$  & 1.5\,$\mu {\rm m}$  & 6.0\,$\mu {\rm m}$
\\
\hline\noalign{\smallskip}
J15582981 &$\mathrm{1224.52}_{-132.28}^{+152.39}$&$\mathrm{334.16}_{-13.52}^{+14.64}$&$\mathrm{171.53}_{-1.98}^{+2.22}$&$\mathrm{33.30}_{-11.88}^{+11.20}$&$\mathrm{0.02}_{-0.02}^{+0.02}$&$\mathrm{18.69}_{-11.62}^{+10.10}$&&$\mathrm{32.86}_{-12.72}^{+12.07}$&$\mathrm{0.02}_{-0.03}^{+0.03}$&$\mathrm{0.05}_{-0.10}^{+0.11}$&&$\mathrm{1.01}_{-0.35}^{+0.34}$&$\mathrm{0.01}_{-0.01}^{+0.01}$&$\mathrm{0.02}_{-0.02}^{+0.02}$&&$\mathrm{0.01}_{-0.02}^{+0.02}$&$\mathrm{0.01}_{-0.01}^{+0.01}$&$\mathrm{13.99}_{-4.63}^{+4.11}$ \\ 
J16035767 &$\mathrm{1497.60}_{-3.84}^{+1.80}$&$\mathrm{273.82}_{-0.40}^{+0.37}$&$\mathrm{96.05}_{-0.19}^{+0.21}$&$\mathrm{47.51}_{-1.72}^{+1.15}$&$\mathrm{0.01}_{-0.00}^{+0.01}$&$\mathrm{0.01}_{-0.01}^{+0.01}$&&$\mathrm{13.11}_{-1.47}^{+1.47}$&$\mathrm{0.01}_{-0.01}^{+0.01}$&$\mathrm{0.01}_{-0.00}^{+0.01}$&&$\mathrm{15.08}_{-0.52}^{+0.48}$&$\mathrm{0.00}_{-0.00}^{+0.00}$&$\mathrm{0.00}_{-0.00}^{+0.00}$&&$\mathrm{6.74}_{-0.26}^{+0.24}$&$\mathrm{17.50}_{-0.64}^{+0.67}$&$\mathrm{0.01}_{-0.02}^{+0.02}$ \\ 
J16062196 &$\mathrm{1000.21}_{-0.16}^{+0.35}$&$\mathrm{144.06}_{-0.29}^{+0.23}$&$\mathrm{158.77}_{-0.17}^{+0.17}$&$\mathrm{0.16}_{-0.00}^{+0.00}$&$\mathrm{0.16}_{-0.00}^{+0.00}$&$\mathrm{0.57}_{-0.47}^{+0.46}$&&$\mathrm{63.53}_{-1.06}^{+1.06}$&$\mathrm{0.17}_{-0.01}^{+0.01}$&$\mathrm{0.17}_{-0.01}^{+0.01}$&&$\mathrm{7.49}_{-0.12}^{+0.11}$&$\mathrm{0.16}_{-0.00}^{+0.00}$&$\mathrm{0.16}_{-0.00}^{+0.00}$&&$\mathrm{9.69}_{-0.16}^{+0.16}$&$\mathrm{17.57}_{-0.25}^{+0.24}$&$\mathrm{0.16}_{-0.00}^{+0.00}$ \\ 
J16064385 &$\mathrm{1492.30}_{-12.35}^{+5.79}$&$\mathrm{278.81}_{-2.51}^{+2.50}$&$\mathrm{398.36}_{-3.04}^{+2.98}$&$\mathrm{0.51}_{-0.01}^{+0.01}$&$\mathrm{0.51}_{-0.01}^{+0.01}$&$\mathrm{0.60}_{-0.08}^{+0.15}$&&$\mathrm{26.17}_{-0.85}^{+0.87}$&$\mathrm{48.50}_{-1.21}^{+1.24}$&$\mathrm{0.86}_{-0.33}^{+0.51}$&&$\mathrm{0.51}_{-0.01}^{+0.01}$&$\mathrm{0.51}_{-0.01}^{+0.01}$&$\mathrm{0.51}_{-0.01}^{+0.01}$&&$\mathrm{6.15}_{-0.15}^{+0.15}$&$\mathrm{14.69}_{-0.41}^{+0.41}$&$\mathrm{0.51}_{-0.01}^{+0.01}$ \\ 
J16075796 &$\mathrm{1163.35}_{-121.47}^{+195.17}$&$\mathrm{617.57}_{-0.62}^{+0.62}$&$\mathrm{132.83}_{-0.18}^{+0.18}$&$\mathrm{36.81}_{-3.29}^{+1.72}$&$\mathrm{0.00}_{-0.00}^{+0.00}$&$\mathrm{0.00}_{-0.00}^{+0.00}$&&$\mathrm{31.66}_{-2.81}^{+1.54}$&$\mathrm{0.01}_{-0.00}^{+0.01}$&$\mathrm{0.00}_{-0.00}^{+0.00}$&&$\mathrm{2.95}_{-0.26}^{+0.14}$&$\mathrm{0.00}_{-0.00}^{+0.00}$&$\mathrm{0.00}_{-0.00}^{+0.00}$&&$\mathrm{0.00}_{-0.00}^{+0.00}$&$\mathrm{0.00}_{-0.00}^{+0.00}$&$\mathrm{28.55}_{-2.56}^{+1.39}$ \\ 
J16111534 &$\mathrm{1165.50}_{-122.39}^{+197.71}$&$\mathrm{574.78}_{-0.52}^{+0.52}$&$\mathrm{235.48}_{-0.44}^{+0.46}$&$\mathrm{46.80}_{-4.17}^{+2.08}$&$\mathrm{0.01}_{-0.00}^{+0.00}$&$\mathrm{0.01}_{-0.00}^{+0.01}$&&$\mathrm{38.39}_{-3.43}^{+1.80}$&$\mathrm{10.79}_{-0.97}^{+0.47}$&$\mathrm{0.01}_{-0.01}^{+0.01}$&&$\mathrm{2.94}_{-0.26}^{+0.13}$&$\mathrm{0.00}_{-0.00}^{+0.00}$&$\mathrm{0.01}_{-0.00}^{+0.00}$&&$\mathrm{1.04}_{-0.09}^{+0.05}$&$\mathrm{0.00}_{-0.00}^{+0.00}$&$\mathrm{0.01}_{-0.00}^{+0.00}$ \\ 
J16120505 &$\mathrm{1499.19}_{-1.34}^{+0.60}$&$\mathrm{229.23}_{-1.35}^{+1.34}$&$\mathrm{385.68}_{-2.56}^{+2.60}$&$\mathrm{1.11}_{-0.03}^{+0.03}$&$\mathrm{1.11}_{-0.03}^{+0.03}$&$\mathrm{1.19}_{-0.07}^{+0.14}$&&$\mathrm{24.77}_{-0.91}^{+0.92}$&$\mathrm{10.75}_{-0.47}^{+0.46}$&$\mathrm{43.39}_{-2.74}^{+2.68}$&&$\mathrm{1.56}_{-0.06}^{+0.06}$&$\mathrm{1.11}_{-0.03}^{+0.03}$&$\mathrm{1.11}_{-0.03}^{+0.03}$&&$\mathrm{3.12}_{-0.11}^{+0.11}$&$\mathrm{9.68}_{-0.37}^{+0.37}$&$\mathrm{1.11}_{-0.03}^{+0.03}$ \\ 
J16123916 &$\mathrm{1499.90}_{-0.16}^{+0.07}$&$\mathrm{185.79}_{-0.35}^{+0.35}$&$\mathrm{309.72}_{-0.41}^{+0.42}$&$\mathrm{0.27}_{-0.00}^{+0.00}$&$\mathrm{0.27}_{-0.00}^{+0.00}$&$\mathrm{0.27}_{-0.00}^{+0.00}$&&$\mathrm{15.73}_{-0.25}^{+0.24}$&$\mathrm{26.84}_{-0.37}^{+0.37}$&$\mathrm{48.11}_{-0.96}^{+0.97}$&&$\mathrm{2.64}_{-0.04}^{+0.04}$&$\mathrm{0.27}_{-0.00}^{+0.00}$&$\mathrm{0.27}_{-0.00}^{+0.00}$&&$\mathrm{1.43}_{-0.02}^{+0.02}$&$\mathrm{3.60}_{-0.07}^{+0.07}$&$\mathrm{0.27}_{-0.00}^{+0.00}$ \\ 
J16141107 &$\mathrm{1174.38}_{-128.19}^{+192.39}$&$\mathrm{475.98}_{-0.15}^{+0.15}$&$\mathrm{133.48}_{-0.23}^{+0.22}$&$\mathrm{0.00}_{-0.00}^{+0.00}$&$\mathrm{0.00}_{-0.00}^{+0.00}$&$\mathrm{25.84}_{-2.00}^{+1.17}$&&$\mathrm{0.00}_{-0.00}^{+0.00}$&$\mathrm{0.00}_{-0.00}^{+0.00}$&$\mathrm{0.01}_{-0.01}^{+0.01}$&&$\mathrm{6.09}_{-0.46}^{+0.21}$&$\mathrm{0.00}_{-0.00}^{+0.00}$&$\mathrm{0.00}_{-0.00}^{+0.00}$&&$\mathrm{10.57}_{-0.79}^{+0.36}$&$\mathrm{15.60}_{-1.16}^{+0.53}$&$\mathrm{0.00}_{-0.00}^{+0.00}$ \\ 
J16142029 &$\mathrm{1180.57}_{-131.82}^{+189.78}$&$\mathrm{455.32}_{-0.14}^{+0.14}$&$\mathrm{407.60}_{-0.36}^{+0.37}$&$\mathrm{35.26}_{-0.22}^{+0.21}$&$\mathrm{0.11}_{-0.00}^{+0.00}$&$\mathrm{0.11}_{-0.01}^{+0.01}$&&$\mathrm{0.11}_{-0.00}^{+0.00}$&$\mathrm{58.04}_{-0.34}^{+0.34}$&$\mathrm{0.11}_{-0.00}^{+0.01}$&&$\mathrm{2.07}_{-0.01}^{+0.01}$&$\mathrm{0.11}_{-0.00}^{+0.00}$&$\mathrm{0.11}_{-0.00}^{+0.00}$&&$\mathrm{1.27}_{-0.02}^{+0.02}$&$\mathrm{2.59}_{-0.04}^{+0.04}$&$\mathrm{0.11}_{-0.00}^{+0.00}$ \\ 
J16153220 &$\mathrm{1183.29}_{-134.05}^{+195.32}$&$\mathrm{392.90}_{-0.23}^{+0.24}$&$\mathrm{240.20}_{-0.61}^{+0.62}$&$\mathrm{61.75}_{-3.86}^{+2.30}$&$\mathrm{0.02}_{-0.00}^{+0.00}$&$\mathrm{0.03}_{-0.02}^{+0.02}$&&$\mathrm{22.20}_{-1.43}^{+0.99}$&$\mathrm{13.49}_{-0.85}^{+0.47}$&$\mathrm{0.03}_{-0.01}^{+0.02}$&&$\mathrm{1.02}_{-0.06}^{+0.04}$&$\mathrm{0.02}_{-0.00}^{+0.00}$&$\mathrm{0.02}_{-0.00}^{+0.00}$&&$\mathrm{0.78}_{-0.05}^{+0.03}$&$\mathrm{0.63}_{-0.07}^{+0.06}$&$\mathrm{0.02}_{-0.01}^{+0.01}$ \\ 
J16153456 &$\mathrm{1374.33}_{-20.96}^{+23.23}$&$\mathrm{998.49}_{-2.53}^{+1.13}$&$\mathrm{401.10}_{-0.73}^{+0.73}$&$\mathrm{62.12}_{-4.93}^{+5.19}$&$\mathrm{0.01}_{-0.00}^{+0.00}$&$\mathrm{0.01}_{-0.00}^{+0.00}$&&$\mathrm{9.26}_{-0.74}^{+0.76}$&$\mathrm{23.22}_{-1.86}^{+1.95}$&$\mathrm{0.01}_{-0.00}^{+0.00}$&&$\mathrm{2.11}_{-0.17}^{+0.18}$&$\mathrm{0.01}_{-0.00}^{+0.00}$&$\mathrm{0.01}_{-0.00}^{+0.00}$&&$\mathrm{0.44}_{-0.04}^{+0.04}$&$\mathrm{2.82}_{-0.23}^{+0.23}$&$\mathrm{0.01}_{-0.00}^{+0.00}$ \\ 
ScoPMS 52 &$\mathrm{1257.98}_{-172.66}^{+165.17}$&$\mathrm{309.34}_{-26.26}^{+36.78}$&$\mathrm{520.33}_{-43.79}^{+43.29}$&$\mathrm{1.07}_{-0.23}^{+0.35}$&$\mathrm{1.38}_{-0.50}^{+0.77}$&$\mathrm{2.63}_{-2.06}^{+2.69}$&&$\mathrm{34.97}_{-12.67}^{+13.83}$&$\mathrm{29.62}_{-10.48}^{+10.46}$&$\mathrm{4.38}_{-4.87}^{+6.10}$&&$\mathrm{5.77}_{-1.88}^{+2.07}$&$\mathrm{1.08}_{-0.24}^{+0.36}$&$\mathrm{1.44}_{-0.56}^{+0.85}$&&$\mathrm{8.09}_{-2.82}^{+3.02}$&$\mathrm{7.94}_{-4.04}^{+3.65}$&$\mathrm{1.64}_{-0.77}^{+1.14}$ \\ 
\hline
\end{tabular}
}
}
\end{table*}

\begin{table*}
\caption{Retrieved temperature and mass fractions of each dust species for Taurus objects.}
\fontsize{40}{90}\selectfont
\label{tab:taurusmfrac}
\rotatebox{0}{%
\resizebox{\textwidth}{!}{
\begin{tabular}{ccccccccccccccccccccccc}
\hline
 \multirow{2}{*}{Object}
&\multirow{2}{*}{$T_{\rm rim}$ (K)} 
&\multirow{2}{*}{$T_{\rm cont}$ (K)}  
&\multirow{2}{*}{$T_{\rm dust}$ (K)}  
&\multicolumn{3}{c}{Olivine (\%)}&
&\multicolumn{3}{c}{Pyroxene (\%)}&
&\multicolumn{3}{c}{Forsterite (\%)}&
&\multicolumn{3}{c}{Enstatite (\%)}\\
\cline{5-7}\cline{9-11}\cline{13-15}\cline{17-19}&&&
&0.1\,$\mu {\rm m}$  & 2.0\,$\mu {\rm m}$  & 5.0\,$\mu {\rm m}$&
&0.1\,$\mu {\rm m}$  & 2.0\,$\mu {\rm m}$  & 5.0\,$\mu {\rm m}$&
&0.1\,$\mu {\rm m}$  & 2.0\,$\mu {\rm m}$  & 5.0\,$\mu {\rm m}$&
&0.1\,$\mu {\rm m}$  & 2.0\,$\mu {\rm m}$  & 5.0\,$\mu {\rm m}$
\\
\hline\noalign{\smallskip} 
AA Tau &$\mathrm{1183.15}_{-133.88}^{+185.67}$&$\mathrm{331.59}_{-35.58}^{+45.31}$&$\mathrm{279.94}_{-28.29}^{+24.40}$&$\mathrm{22.73}_{-14.52}^{+14.64}$&$\mathrm{0.22}_{-0.24}^{+0.31}$&$\mathrm{0.53}_{-0.86}^{+1.00}$&&$\mathrm{30.88}_{-23.25}^{+28.11}$&$\mathrm{0.35}_{-0.48}^{+0.50}$&$\mathrm{0.48}_{-0.76}^{+0.87}$&&$\mathrm{0.14}_{-0.11}^{+0.12}$&$\mathrm{0.14}_{-0.10}^{+0.14}$&$\mathrm{0.22}_{-0.24}^{+0.30}$&&$\mathrm{0.19}_{-0.18}^{+0.22}$&$\mathrm{4.71}_{-6.61}^{+3.47}$&$\mathrm{39.41}_{-31.24}^{+33.29}$ \\ 
BP Tau &$\mathrm{1184.00}_{-134.19}^{+188.91}$&$\mathrm{392.99}_{-51.96}^{+38.09}$&$\mathrm{242.54}_{-18.56}^{+35.79}$&$\mathrm{38.33}_{-45.54}^{+30.27}$&$\mathrm{0.13}_{-0.21}^{+0.23}$&$\mathrm{0.37}_{-0.82}^{+0.86}$&&$\mathrm{58.59}_{-82.25}^{+55.46}$&$\mathrm{0.16}_{-0.29}^{+0.31}$&$\mathrm{0.31}_{-0.66}^{+0.71}$&&$\mathrm{0.75}_{-1.04}^{+0.72}$&$\mathrm{0.09}_{-0.12}^{+0.13}$&$\mathrm{0.13}_{-0.22}^{+0.24}$&&$\mathrm{0.36}_{-0.76}^{+0.45}$&$\mathrm{0.47}_{-1.11}^{+0.73}$&$\mathrm{0.31}_{-0.68}^{+0.70}$ \\ 
CI Tau &$\mathrm{1170.82}_{-121.88}^{+187.27}$&$\mathrm{497.41}_{-79.95}^{+53.74}$&$\mathrm{247.75}_{-21.35}^{+25.37}$&$\mathrm{36.16}_{-31.10}^{+30.84}$&$\mathrm{0.23}_{-0.45}^{+0.48}$&$\mathrm{0.95}_{-2.69}^{+2.76}$&&$\mathrm{33.14}_{-36.86}^{+36.97}$&$\mathrm{16.16}_{-14.11}^{+12.78}$&$\mathrm{0.45}_{-1.08}^{+1.20}$&&$\mathrm{1.79}_{-1.52}^{+1.49}$&$\mathrm{0.08}_{-0.10}^{+0.13}$&$\mathrm{0.18}_{-0.32}^{+0.34}$&&$\mathrm{1.41}_{-1.67}^{+1.42}$&$\mathrm{5.49}_{-8.03}^{+5.76}$&$\mathrm{3.98}_{-13.25}^{+9.67}$ \\ 
CW Tau &$\mathrm{1185.95}_{-121.13}^{+162.04}$&$\mathrm{357.65}_{-16.66}^{+17.13}$&$\mathrm{202.85}_{-10.62}^{+11.24}$&$\mathrm{22.75}_{-10.87}^{+11.26}$&$\mathrm{0.09}_{-0.09}^{+0.10}$&$\mathrm{0.15}_{-0.21}^{+0.27}$&&$\mathrm{21.16}_{-13.90}^{+13.89}$&$\mathrm{0.12}_{-0.14}^{+0.16}$&$\mathrm{0.12}_{-0.15}^{+0.18}$&&$\mathrm{3.10}_{-1.28}^{+1.31}$&$\mathrm{0.04}_{-0.02}^{+0.03}$&$\mathrm{0.08}_{-0.07}^{+0.09}$&&$\mathrm{4.87}_{-2.01}^{+2.05}$&$\mathrm{0.11}_{-0.13}^{+0.14}$&$\mathrm{47.42}_{-18.50}^{+18.08}$ \\ 
CX Tau &$\mathrm{1185.42}_{-136.18}^{+182.41}$&$\mathrm{388.86}_{-33.21}^{+21.66}$&$\mathrm{178.95}_{-5.98}^{+6.58}$&$\mathrm{0.15}_{-0.33}^{+0.33}$&$\mathrm{0.10}_{-0.19}^{+0.20}$&$\mathrm{0.15}_{-0.35}^{+0.40}$&&$\mathrm{93.25}_{-52.38}^{+28.11}$&$\mathrm{0.11}_{-0.22}^{+0.25}$&$\mathrm{0.14}_{-0.30}^{+0.32}$&&$\mathrm{0.69}_{-0.60}^{+0.37}$&$\mathrm{0.04}_{-0.05}^{+0.06}$&$\mathrm{0.06}_{-0.09}^{+0.11}$&&$\mathrm{4.24}_{-2.51}^{+1.65}$&$\mathrm{0.89}_{-3.09}^{+1.30}$&$\mathrm{0.20}_{-0.47}^{+0.50}$ \\ 
CY Tau &$\mathrm{1393.51}_{-132.99}^{+77.20}$&$\mathrm{261.69}_{-19.76}^{+24.87}$&$\mathrm{503.95}_{-49.24}^{+58.70}$&$\mathrm{3.17}_{-0.66}^{+1.01}$&$\mathrm{3.38}_{-0.80}^{+1.28}$&$\mathrm{6.95}_{-4.81}^{+6.54}$&&$\mathrm{43.11}_{-13.18}^{+14.90}$&$\mathrm{3.44}_{-0.87}^{+1.37}$&$\mathrm{7.07}_{-5.11}^{+6.81}$&&$\mathrm{3.41}_{-0.78}^{+1.03}$&$\mathrm{3.20}_{-0.66}^{+1.05}$&$\mathrm{4.05}_{-1.41}^{+2.25}$&&$\mathrm{12.19}_{-3.42}^{+4.02}$&$\mathrm{5.12}_{-2.42}^{+2.85}$&$\mathrm{4.91}_{-2.33}^{+3.41}$ \\ 
DE Tau &$\mathrm{1188.28}_{-129.96}^{+173.84}$&$\mathrm{275.10}_{-28.99}^{+40.26}$&$\mathrm{220.76}_{-14.10}^{+17.26}$&$\mathrm{40.86}_{-25.17}^{+25.82}$&$\mathrm{0.14}_{-0.16}^{+0.19}$&$\mathrm{0.34}_{-0.57}^{+0.65}$&&$\mathrm{39.50}_{-30.23}^{+29.50}$&$\mathrm{0.20}_{-0.28}^{+0.30}$&$\mathrm{0.33}_{-0.58}^{+0.65}$&&$\mathrm{3.16}_{-1.79}^{+1.87}$&$\mathrm{0.10}_{-0.09}^{+0.12}$&$\mathrm{0.14}_{-0.17}^{+0.21}$&&$\mathrm{0.66}_{-1.08}^{+0.58}$&$\mathrm{14.07}_{-7.73}^{+7.61}$&$\mathrm{0.51}_{-1.01}^{+1.00}$ \\ 
DG Tau &$\mathrm{1240.51}_{-164.34}^{+141.24}$&$\mathrm{189.25}_{-14.74}^{+17.03}$&$\mathrm{121.49}_{-4.46}^{+6.10}$&$\mathrm{0.57}_{-1.48}^{+1.49}$&$\mathrm{0.38}_{-0.99}^{+0.83}$&$\mathrm{0.11}_{-0.24}^{+0.38}$&&$\mathrm{96.15}_{-116.07}^{+54.76}$&$\mathrm{0.43}_{-1.05}^{+0.92}$&$\mathrm{0.19}_{-0.45}^{+0.68}$&&$\mathrm{0.26}_{-0.58}^{+0.43}$&$\mathrm{0.33}_{-0.84}^{+0.64}$&$\mathrm{0.22}_{-0.54}^{+0.58}$&&$\mathrm{0.74}_{-1.94}^{+1.58}$&$\mathrm{0.16}_{-0.36}^{+0.40}$&$\mathrm{0.45}_{-1.23}^{+1.19}$ \\ 
DL Tau &$\mathrm{1170.92}_{-128.01}^{+199.50}$&$\mathrm{437.33}_{-62.40}^{+17.68}$&$\mathrm{155.09}_{-11.55}^{+21.68}$&$\mathrm{0.17}_{-0.37}^{+0.37}$&$\mathrm{0.12}_{-0.24}^{+0.27}$&$\mathrm{1.20}_{-4.02}^{+3.76}$&&$\mathrm{77.70}_{-124.14}^{+75.44}$&$\mathrm{0.21}_{-0.49}^{+0.49}$&$\mathrm{0.48}_{-1.29}^{+1.35}$&&$\mathrm{17.71}_{-27.03}^{+15.70}$&$\mathrm{0.08}_{-0.14}^{+0.16}$&$\mathrm{0.17}_{-0.35}^{+0.36}$&&$\mathrm{1.59}_{-4.65}^{+3.29}$&$\mathrm{0.25}_{-0.60}^{+0.57}$&$\mathrm{0.32}_{-0.78}^{+0.74}$ \\ 
DM Tau &$\mathrm{1197.11}_{-139.94}^{+184.99}$&$\mathrm{218.72}_{-5.08}^{+4.69}$&$\mathrm{145.52}_{-3.40}^{+4.41}$&$\mathrm{41.79}_{-25.95}^{+16.87}$&$\mathrm{0.08}_{-0.18}^{+0.20}$&$\mathrm{0.15}_{-0.38}^{+0.38}$&&$\mathrm{44.92}_{-32.30}^{+21.31}$&$\mathrm{11.16}_{-7.87}^{+5.06}$&$\mathrm{0.44}_{-1.21}^{+0.95}$&&$\mathrm{0.95}_{-0.57}^{+0.38}$&$\mathrm{0.06}_{-0.12}^{+0.12}$&$\mathrm{0.09}_{-0.18}^{+0.19}$&&$\mathrm{0.06}_{-0.11}^{+0.10}$&$\mathrm{0.09}_{-0.21}^{+0.20}$&$\mathrm{0.19}_{-0.52}^{+0.51}$ \\ 
DN Tau &$\mathrm{1266.51}_{-179.37}^{+156.25}$&$\mathrm{214.20}_{-12.33}^{+16.29}$&$\mathrm{244.78}_{-12.15}^{+12.86}$&$\mathrm{0.50}_{-0.21}^{+0.29}$&$\mathrm{0.70}_{-0.44}^{+0.60}$&$\mathrm{1.95}_{-2.46}^{+2.88}$&&$\mathrm{32.84}_{-14.54}^{+15.74}$&$\mathrm{50.99}_{-18.79}^{+19.70}$&$\mathrm{1.74}_{-2.12}^{+2.64}$&&$\mathrm{0.35}_{-0.09}^{+0.11}$&$\mathrm{0.44}_{-0.15}^{+0.22}$&$\mathrm{0.67}_{-0.40}^{+0.55}$&&$\mathrm{0.62}_{-0.33}^{+0.42}$&$\mathrm{3.25}_{-4.36}^{+2.97}$&$\mathrm{5.96}_{-10.66}^{+8.20}$ \\ 
DO Tau &$\mathrm{1178.64}_{-130.60}^{+196.80}$&$\mathrm{343.61}_{-33.53}^{+34.72}$&$\mathrm{157.59}_{-6.60}^{+10.30}$&$\mathrm{0.22}_{-0.53}^{+0.55}$&$\mathrm{0.19}_{-0.44}^{+0.44}$&$\mathrm{0.34}_{-0.93}^{+1.02}$&&$\mathrm{74.43}_{-74.02}^{+45.94}$&$\mathrm{0.51}_{-1.48}^{+1.26}$&$\mathrm{0.33}_{-0.92}^{+0.99}$&&$\mathrm{5.12}_{-4.43}^{+2.98}$&$\mathrm{0.08}_{-0.14}^{+0.15}$&$\mathrm{0.10}_{-0.20}^{+0.23}$&&$\mathrm{3.60}_{-4.71}^{+3.11}$&$\mathrm{13.62}_{-15.92}^{+9.48}$&$\mathrm{1.45}_{-5.27}^{+4.33}$ \\ 
DR Tau &$\mathrm{1183.21}_{-137.77}^{+196.73}$&$\mathrm{450.51}_{-43.31}^{+33.31}$&$\mathrm{293.63}_{-29.00}^{+23.02}$&$\mathrm{79.41}_{-52.93}^{+58.80}$&$\mathrm{0.88}_{-1.37}^{+1.50}$&$\mathrm{1.80}_{-3.57}^{+3.90}$&&$\mathrm{2.46}_{-5.49}^{+5.17}$&$\mathrm{1.22}_{-2.17}^{+2.16}$&$\mathrm{2.24}_{-4.82}^{+4.99}$&&$\mathrm{0.29}_{-0.25}^{+0.26}$&$\mathrm{0.46}_{-0.54}^{+0.62}$&$\mathrm{0.79}_{-1.23}^{+1.27}$&&$\mathrm{0.87}_{-1.29}^{+0.96}$&$\mathrm{4.00}_{-7.33}^{+3.68}$&$\mathrm{5.59}_{-15.48}^{+9.86}$ \\ 
DS Tau &$\mathrm{1179.91}_{-129.57}^{+191.23}$&$\mathrm{465.86}_{-53.43}^{+41.75}$&$\mathrm{240.64}_{-13.52}^{+15.45}$&$\mathrm{45.06}_{-27.53}^{+27.06}$&$\mathrm{0.07}_{-0.10}^{+0.13}$&$\mathrm{0.21}_{-0.45}^{+0.53}$&&$\mathrm{42.22}_{-29.95}^{+29.88}$&$\mathrm{0.13}_{-0.23}^{+0.25}$&$\mathrm{0.21}_{-0.44}^{+0.48}$&&$\mathrm{2.04}_{-1.16}^{+1.14}$&$\mathrm{0.05}_{-0.05}^{+0.07}$&$\mathrm{0.08}_{-0.13}^{+0.15}$&&$\mathrm{0.48}_{-1.00}^{+0.45}$&$\mathrm{9.24}_{-5.24}^{+4.73}$&$\mathrm{0.21}_{-0.45}^{+0.52}$ \\ 
FM Tau &$\mathrm{1163.78}_{-118.85}^{+190.49}$&$\mathrm{375.20}_{-33.98}^{+25.03}$&$\mathrm{239.21}_{-9.16}^{+11.28}$&$\mathrm{76.68}_{-34.77}^{+33.61}$&$\mathrm{0.12}_{-0.17}^{+0.19}$&$\mathrm{0.39}_{-0.83}^{+0.83}$&&$\mathrm{20.06}_{-11.94}^{+10.78}$&$\mathrm{0.16}_{-0.24}^{+0.26}$&$\mathrm{0.30}_{-0.60}^{+0.62}$&&$\mathrm{0.88}_{-0.40}^{+0.36}$&$\mathrm{0.09}_{-0.11}^{+0.13}$&$\mathrm{0.13}_{-0.20}^{+0.22}$&&$\mathrm{0.78}_{-1.06}^{+0.51}$&$\mathrm{0.17}_{-0.26}^{+0.25}$&$\mathrm{0.24}_{-0.44}^{+0.48}$ \\ 
FN Tau &$\mathrm{1238.90}_{-164.02}^{+179.39}$&$\mathrm{293.66}_{-34.22}^{+31.95}$&$\mathrm{159.44}_{-6.03}^{+5.97}$&$\mathrm{0.14}_{-0.33}^{+0.33}$&$\mathrm{0.06}_{-0.11}^{+0.15}$&$\mathrm{0.14}_{-0.33}^{+0.38}$&&$\mathrm{82.95}_{-33.83}^{+17.60}$&$\mathrm{0.12}_{-0.26}^{+0.30}$&$\mathrm{0.15}_{-0.37}^{+0.41}$&&$\mathrm{0.19}_{-0.44}^{+0.23}$&$\mathrm{0.05}_{-0.08}^{+0.09}$&$\mathrm{0.09}_{-0.19}^{+0.20}$&&$\mathrm{7.57}_{-3.34}^{+2.44}$&$\mathrm{8.39}_{-6.36}^{+3.80}$&$\mathrm{0.14}_{-0.32}^{+0.34}$ \\ 
FP Tau &$\mathrm{1236.46}_{-151.19}^{+156.86}$&$\mathrm{217.65}_{-20.17}^{+25.25}$&$\mathrm{167.29}_{-5.36}^{+7.71}$&$\mathrm{0.32}_{-0.70}^{+0.64}$&$\mathrm{0.08}_{-0.11}^{+0.12}$&$\mathrm{0.28}_{-0.61}^{+0.71}$&&$\mathrm{96.36}_{-84.33}^{+69.61}$&$\mathrm{0.07}_{-0.10}^{+0.12}$&$\mathrm{0.19}_{-0.36}^{+0.42}$&&$\mathrm{2.07}_{-1.71}^{+1.44}$&$\mathrm{0.06}_{-0.07}^{+0.09}$&$\mathrm{0.10}_{-0.16}^{+0.19}$&&$\mathrm{0.06}_{-0.07}^{+0.08}$&$\mathrm{0.21}_{-0.43}^{+0.38}$&$\mathrm{0.20}_{-0.39}^{+0.40}$ \\ 
FT Tau &$\mathrm{1192.33}_{-140.82}^{+198.60}$&$\mathrm{381.47}_{-28.90}^{+14.84}$&$\mathrm{300.16}_{-28.60}^{+29.81}$&$\mathrm{35.58}_{-23.64}^{+23.54}$&$\mathrm{0.52}_{-0.62}^{+0.76}$&$\mathrm{1.02}_{-1.65}^{+1.93}$&&$\mathrm{33.41}_{-27.96}^{+28.66}$&$\mathrm{7.80}_{-17.20}^{+6.16}$&$\mathrm{1.22}_{-2.14}^{+2.30}$&&$\mathrm{3.63}_{-2.39}^{+2.35}$&$\mathrm{0.32}_{-0.29}^{+0.36}$&$\mathrm{0.42}_{-0.45}^{+0.57}$&&$\mathrm{0.30}_{-0.26}^{+0.31}$&$\mathrm{10.90}_{-9.22}^{+7.79}$&$\mathrm{4.89}_{-12.99}^{+9.58}$ \\ 
FZ Tau &$\mathrm{1174.12}_{-128.91}^{+201.36}$&$\mathrm{492.64}_{-35.16}^{+22.96}$&$\mathrm{146.34}_{-5.45}^{+5.53}$&$\mathrm{0.06}_{-0.14}^{+0.16}$&$\mathrm{0.06}_{-0.13}^{+0.15}$&$\mathrm{0.17}_{-0.50}^{+0.53}$&&$\mathrm{0.11}_{-0.30}^{+0.31}$&$\mathrm{0.09}_{-0.23}^{+0.25}$&$\mathrm{0.12}_{-0.31}^{+0.35}$&&$\mathrm{14.40}_{-4.79}^{+3.00}$&$\mathrm{0.04}_{-0.08}^{+0.08}$&$\mathrm{0.08}_{-0.17}^{+0.18}$&&$\mathrm{23.37}_{-8.54}^{+5.68}$&$\mathrm{19.30}_{-9.68}^{+6.48}$&$\mathrm{0.19}_{-0.56}^{+0.58}$ \\ 
GI Tau &$\mathrm{1137.31}_{-101.89}^{+191.79}$&$\mathrm{473.03}_{-53.39}^{+68.86}$&$\mathrm{263.29}_{-20.54}^{+29.76}$&$\mathrm{41.20}_{-43.00}^{+39.15}$&$\mathrm{0.17}_{-0.35}^{+0.38}$&$\mathrm{0.94}_{-2.64}^{+2.44}$&&$\mathrm{27.33}_{-34.69}^{+29.66}$&$\mathrm{0.94}_{-2.54}^{+1.61}$&$\mathrm{27.18}_{-67.13}^{+38.73}$&&$\mathrm{1.00}_{-1.17}^{+0.99}$&$\mathrm{0.11}_{-0.20}^{+0.18}$&$\mathrm{0.15}_{-0.29}^{+0.30}$&&$\mathrm{0.09}_{-0.15}^{+0.13}$&$\mathrm{0.33}_{-0.82}^{+0.65}$&$\mathrm{0.57}_{-1.55}^{+1.35}$ \\ 
GK Tau &$\mathrm{1150.92}_{-111.51}^{+181.05}$&$\mathrm{423.93}_{-40.39}^{+25.71}$&$\mathrm{262.70}_{-12.85}^{+15.71}$&$\mathrm{44.72}_{-24.70}^{+23.72}$&$\mathrm{0.13}_{-0.19}^{+0.21}$&$\mathrm{0.30}_{-0.64}^{+0.69}$&&$\mathrm{41.91}_{-26.12}^{+25.56}$&$\mathrm{0.25}_{-0.49}^{+0.52}$&$\mathrm{0.47}_{-1.18}^{+1.24}$&&$\mathrm{3.00}_{-1.59}^{+1.55}$&$\mathrm{0.07}_{-0.09}^{+0.10}$&$\mathrm{0.13}_{-0.20}^{+0.22}$&&$\mathrm{0.26}_{-0.47}^{+0.27}$&$\mathrm{8.44}_{-4.42}^{+3.97}$&$\mathrm{0.32}_{-0.69}^{+0.70}$ \\ 
GM Aur &$\mathrm{1192.77}_{-139.60}^{+192.18}$&$\mathrm{414.31}_{-42.35}^{+28.05}$&$\mathrm{294.27}_{-5.89}^{+6.38}$&$\mathrm{95.61}_{-43.97}^{+43.76}$&$\mathrm{0.10}_{-0.14}^{+0.16}$&$\mathrm{0.68}_{-1.83}^{+1.78}$&&$\mathrm{0.12}_{-0.18}^{+0.20}$&$\mathrm{0.12}_{-0.19}^{+0.20}$&$\mathrm{0.28}_{-0.59}^{+0.66}$&&$\mathrm{1.15}_{-0.52}^{+0.52}$&$\mathrm{0.06}_{-0.06}^{+0.08}$&$\mathrm{0.08}_{-0.11}^{+0.13}$&&$\mathrm{0.03}_{-0.02}^{+0.03}$&$\mathrm{0.12}_{-0.17}^{+0.18}$&$\mathrm{1.66}_{-5.09}^{+2.40}$ \\ 
GO Tau &$\mathrm{1192.05}_{-138.61}^{+193.73}$&$\mathrm{420.41}_{-45.12}^{+27.77}$&$\mathrm{273.52}_{-18.59}^{+22.27}$&$\mathrm{0.21}_{-0.25}^{+0.31}$&$\mathrm{0.18}_{-0.19}^{+0.24}$&$\mathrm{0.62}_{-1.18}^{+1.36}$&&$\mathrm{44.18}_{-29.82}^{+28.38}$&$\mathrm{0.23}_{-0.28}^{+0.35}$&$\mathrm{0.43}_{-0.72}^{+0.86}$&&$\mathrm{4.77}_{-2.52}^{+2.48}$&$\mathrm{0.12}_{-0.11}^{+0.14}$&$\mathrm{0.19}_{-0.22}^{+0.27}$&&$\mathrm{0.08}_{-0.05}^{+0.07}$&$\mathrm{48.35}_{-25.39}^{+24.61}$&$\mathrm{0.63}_{-1.22}^{+1.34}$ \\ 
HO Tau &$\mathrm{1192.28}_{-144.19}^{+196.58}$&$\mathrm{447.48}_{-49.20}^{+30.67}$&$\mathrm{205.73}_{-6.92}^{+8.68}$&$\mathrm{59.77}_{-34.71}^{+24.37}$&$\mathrm{0.09}_{-0.19}^{+0.20}$&$\mathrm{0.18}_{-0.45}^{+0.55}$&&$\mathrm{34.28}_{-22.90}^{+16.07}$&$\mathrm{0.14}_{-0.33}^{+0.31}$&$\mathrm{0.18}_{-0.45}^{+0.49}$&&$\mathrm{1.39}_{-0.75}^{+0.56}$&$\mathrm{0.50}_{-1.55}^{+0.75}$&$\mathrm{0.17}_{-0.43}^{+0.47}$&&$\mathrm{3.10}_{-1.85}^{+1.39}$&$\mathrm{0.06}_{-0.09}^{+0.11}$&$\mathrm{0.14}_{-0.32}^{+0.36}$ \\ 
HP Tau &$\mathrm{1190.04}_{-138.55}^{+190.88}$&$\mathrm{373.90}_{-16.73}^{+9.81}$&$\mathrm{230.02}_{-9.65}^{+13.56}$&$\mathrm{56.76}_{-28.71}^{+24.84}$&$\mathrm{0.15}_{-0.19}^{+0.23}$&$\mathrm{0.31}_{-0.58}^{+0.64}$&&$\mathrm{38.82}_{-24.16}^{+19.89}$&$\mathrm{0.22}_{-0.34}^{+0.39}$&$\mathrm{0.30}_{-0.53}^{+0.59}$&&$\mathrm{2.05}_{-0.97}^{+0.85}$&$\mathrm{0.08}_{-0.07}^{+0.09}$&$\mathrm{0.13}_{-0.15}^{+0.19}$&&$\mathrm{0.07}_{-0.06}^{+0.07}$&$\mathrm{0.52}_{-1.07}^{+0.69}$&$\mathrm{0.60}_{-1.38}^{+1.28}$ \\ 
IP Tau &$\mathrm{1137.42}_{-100.54}^{+189.86}$&$\mathrm{486.24}_{-39.59}^{+36.69}$&$\mathrm{297.58}_{-14.43}^{+18.03}$&$\mathrm{63.69}_{-36.27}^{+39.38}$&$\mathrm{0.12}_{-0.18}^{+0.22}$&$\mathrm{0.41}_{-0.90}^{+0.98}$&&$\mathrm{26.14}_{-17.48}^{+18.80}$&$\mathrm{3.31}_{-9.04}^{+2.66}$&$\mathrm{3.12}_{-9.79}^{+6.72}$&&$\mathrm{2.04}_{-1.12}^{+1.23}$&$\mathrm{0.07}_{-0.09}^{+0.10}$&$\mathrm{0.10}_{-0.13}^{+0.16}$&&$\mathrm{0.08}_{-0.10}^{+0.10}$&$\mathrm{0.62}_{-1.57}^{+0.91}$&$\mathrm{0.31}_{-0.65}^{+0.71}$ \\ 
IQ Tau &$\mathrm{1137.76}_{-100.59}^{+164.18}$&$\mathrm{521.31}_{-70.81}^{+59.09}$&$\mathrm{260.96}_{-26.05}^{+14.29}$&$\mathrm{54.64}_{-35.60}^{+48.13}$&$\mathrm{0.52}_{-1.11}^{+0.97}$&$\mathrm{4.69}_{-15.89}^{+10.73}$&&$\mathrm{1.65}_{-5.23}^{+4.37}$&$\mathrm{32.47}_{-20.12}^{+23.10}$&$\mathrm{0.96}_{-2.57}^{+2.33}$&&$\mathrm{1.97}_{-1.24}^{+1.71}$&$\mathrm{0.10}_{-0.13}^{+0.15}$&$\mathrm{0.16}_{-0.28}^{+0.30}$&&$\mathrm{0.07}_{-0.08}^{+0.10}$&$\mathrm{0.22}_{-0.40}^{+0.36}$&$\mathrm{2.55}_{-8.17}^{+5.08}$ \\ 
LkCa 15 &$\mathrm{1155.30}_{-116.12}^{+194.07}$&$\mathrm{521.14}_{-71.10}^{+46.69}$&$\mathrm{407.37}_{-20.31}^{+21.92}$&$\mathrm{90.85}_{-39.47}^{+39.35}$&$\mathrm{0.09}_{-0.10}^{+0.12}$&$\mathrm{0.20}_{-0.31}^{+0.37}$&&$\mathrm{2.54}_{-3.90}^{+2.09}$&$\mathrm{0.22}_{-0.35}^{+0.37}$&$\mathrm{0.19}_{-0.29}^{+0.33}$&&$\mathrm{1.62}_{-0.67}^{+0.65}$&$\mathrm{0.06}_{-0.05}^{+0.07}$&$\mathrm{0.10}_{-0.11}^{+0.13}$&&$\mathrm{0.11}_{-0.13}^{+0.13}$&$\mathrm{3.79}_{-1.82}^{+1.53}$&$\mathrm{0.22}_{-0.35}^{+0.40}$ \\ 
V410 X-ray 1 &$\mathrm{1159.43}_{-119.32}^{+193.07}$&$\mathrm{428.25}_{-78.20}^{+70.54}$&$\mathrm{205.87}_{-7.96}^{+8.90}$&$\mathrm{42.75}_{-21.07}^{+13.55}$&$\mathrm{0.08}_{-0.17}^{+0.18}$&$\mathrm{0.09}_{-0.20}^{+0.23}$&&$\mathrm{52.32}_{-29.16}^{+16.99}$&$\mathrm{0.46}_{-1.54}^{+1.01}$&$\mathrm{0.17}_{-0.45}^{+0.43}$&&$\mathrm{2.28}_{-1.11}^{+0.72}$&$\mathrm{0.10}_{-0.24}^{+0.20}$&$\mathrm{0.09}_{-0.19}^{+0.20}$&&$\mathrm{0.04}_{-0.06}^{+0.06}$&$\mathrm{1.21}_{-4.25}^{+1.27}$&$\mathrm{0.41}_{-1.33}^{+1.10}$ \\ 
V836 Tau &$\mathrm{1174.25}_{-128.28}^{+192.49}$&$\mathrm{427.21}_{-57.96}^{+51.58}$&$\mathrm{341.12}_{-27.51}^{+26.95}$&$\mathrm{17.42}_{-8.24}^{+8.62}$&$\mathrm{0.21}_{-0.22}^{+0.26}$&$\mathrm{0.55}_{-0.90}^{+1.04}$&&$\mathrm{48.27}_{-26.34}^{+28.99}$&$\mathrm{0.27}_{-0.30}^{+0.36}$&$\mathrm{0.35}_{-0.48}^{+0.59}$&&$\mathrm{3.83}_{-1.71}^{+1.82}$&$\mathrm{0.11}_{-0.07}^{+0.10}$&$\mathrm{0.18}_{-0.17}^{+0.22}$&&$\mathrm{0.33}_{-0.39}^{+0.32}$&$\mathrm{27.83}_{-12.37}^{+12.98}$&$\mathrm{0.64}_{-1.13}^{+1.19}$ \\ 
ZZ Tau &$\mathrm{1178.14}_{-126.54}^{+190.45}$&$\mathrm{442.61}_{-24.06}^{+17.58}$&$\mathrm{152.02}_{-6.56}^{+8.91}$&$\mathrm{30.97}_{-21.32}^{+11.45}$&$\mathrm{0.11}_{-0.26}^{+0.27}$&$\mathrm{0.04}_{-0.09}^{+0.11}$&&$\mathrm{43.17}_{-25.97}^{+10.12}$&$\mathrm{10.33}_{-44.58}^{+10.62}$&$\mathrm{0.09}_{-0.22}^{+0.22}$&&$\mathrm{3.67}_{-2.09}^{+1.20}$&$\mathrm{0.05}_{-0.10}^{+0.11}$&$\mathrm{0.06}_{-0.15}^{+0.17}$&&$\mathrm{2.46}_{-3.25}^{+1.44}$&$\mathrm{8.97}_{-14.81}^{+5.22}$&$\mathrm{0.08}_{-0.20}^{+0.28}$ \\ 
\hline
\end{tabular}
}
}
\end{table*}

\end{document}